\documentclass[%
 reprint,
 amsmath,amssymb,
 aps,
]{revtex4-2}

\usepackage{xcolor}

\usepackage{graphicx}% Include figure files
\usepackage{dcolumn}% Align table columns on decimal point
\usepackage{bm}% bold math
\begin{document}

\preprint{APS/123-QED}

\title{Nonequilibrium Hanbury-Brown-Twiss experiment: Theory and application to binary stars}%\\with Forced Linebreak}% Force line breaks with \\

\author{Adrian E. Rubio L\'opez}\email{adrianrubiolopez0102@gmail.com}
\author{Ashwin K. Boddeti}
\author{Fanglin Bao}
\author{Hyunsoo Choi}
\author{Zubin Jacob}\email{zjacob@purdue.edu}
\homepage{http://www.electrodynamics.org}
\affiliation{%
 Birck Nanotechnology Center, School of Electrical and Computer Engineering,\\
 Purdue University, West Lafayette, IN 47907, USA%\\
}%


\date{\today}% It is always \today, today,

\begin{abstract}
Intensity-interferometry based on Hanbury-Brown and Twiss’s seminal experiment for determining the radius of the star Sirius formed the basis for developing the quantum theory of light. To date, the principle of this experiment is used in various forms across different fields of quantum optics, imaging and astronomy. Though, the technique is powerful, it has not been generalized for objects at different temperatures. Here, we address this problem using a generating functional formalism by employing the P-function representation of quantum-thermal light. Specifically, we investigate the photon coincidences of a system of two extended objects at different temperature using this theoretical framework. We show two unique aspects in the second-order quantum coherence function - interference oscillations and a long-baseline asymptotic value that depends on the observation frequency, temperatures and size of both objects. We apply our approach to the case of binary stars and discuss the advantages of measuring these two features in an experiment. In addition to the estimation of the radii of each star and the distance between them, we also show that the present approach is suitable for the estimation of temperatures as well. To this end, we apply it to the practical case of binary stars  Luhman 16 and Spica $\alpha$ Vir. We find that for currently available telescopes, an experimental demonstration is feasible in the near term. Our work contributes to the fundamental understanding of intensity interferometry of quantum-thermal light and can be used as a tool for studying two-body thermal emitters - from binary stars to extended objects.
\end{abstract}

\maketitle

%\tableofcontents

%%%%%%%%%%%%%%%%%%%%%%%%%%%%%
\section{\label{sec:Intro}Introduction}
%%%%%%%%%%%%%%%%%%%%%%%%%%%%%

Hanbury-Brown and Twiss (HBT) in 1956 reported that photons with narrow spectral width coming from the star Sirius have a tendency to arrive as correlated pairs \cite{HBT1956}. This observation turned out to be the most prominent experiment that lead to the development of a quantum mechanical description of photon correlations\cite{Glauber,Sudarshan}. The intensity-interferometry experiments of Hanbury-Brown and Twiss were of central importance in the study photon correlations, and thus  the quantum theory of light that is subject of constant investigation to date across fields from cosmology, nuclear physics to atomic flourescence \cite{Antibunching,Plasma, Baym,Cosmology, DifferentialHBT,Nanoresonator,NonclassicalGW,SimulationsBinary,VonZanthier2021}. The intensity interferometry experiment quantified the intensity correlations (coincidences) of light coming from from a source or a system of sources which is incident on two separate detectors. Similar experiments also allows the classification of sources as--- single-photon, coherent or in-coherent in nature. Despite the low-mode occupancy of thermal light, intensity-interferometry was employed to estimate the angular size of the star Sirius A by Hanbury-Brown and Twiss.% This was a major breakthrough since it was a pristine utilization of high-order statistical properties of photons for measurements on astrophysical objects such as stars. In this sense, intensity-interferometry not only helps study fundamental physics of the quantum nature of light but also provides a very powerful tool for exploring and understanding the universe.

Intensity-interferometry from thermal sources like stars is employed to extract useful astronomical information such as the source (intensity) distribution and size of astronomical objects. More recently, quantum imaging have helped in overcoming the limitations set by diffraction-limited optics in microscopy \cite{thiel2009sub,cui2013quantum,israel2017quantum,classen2017superresolution,tenne2019super,forbes2019super} and also to resolve astronomical sources that otherwise were not resolved by interferometry techniques\cite{Croke,tsang2016quantum,BaoQAI}. All the current techniques to date are limited in terms of the amount of information that can be extracted (for example temperature) and often rely on  other techniques (such as spectroscopy) to complement these measurements. For both applications, i.e. estimating the size and spatial distribution of sources, the temperature distribution of the sources has not been taken into consideration. Furthermore, owing to the fact that in intensity interferometric methods the signal scales quadratically with the mean photon number $\bar{n}  = ({\rm Exp}[\hbar\omega/(k_{\rm B}T)] - 1)^{-1}$ it is important to incorporate and consider the temperature distributions of the astronomical sources. Therefore, incorporating the individual temperature of the objects of interest, one can fully characterize the system (temperature, distance between them, and angular sizes) without relying on other complementary measurements. In this work, we aim to address the scenario when the objects of interest are at two arbitrary distinct temperatures as compared to the conventional HBT experiment which only considers a single object at a uniform temperature. We develop a theoretical framework and also provide an analysis of the measurement strategies to adopt depending on the experimental conditions available which include the relative motion of a pair of bounded objects.

\begin{figure}[!ht]
\includegraphics[width=\linewidth]{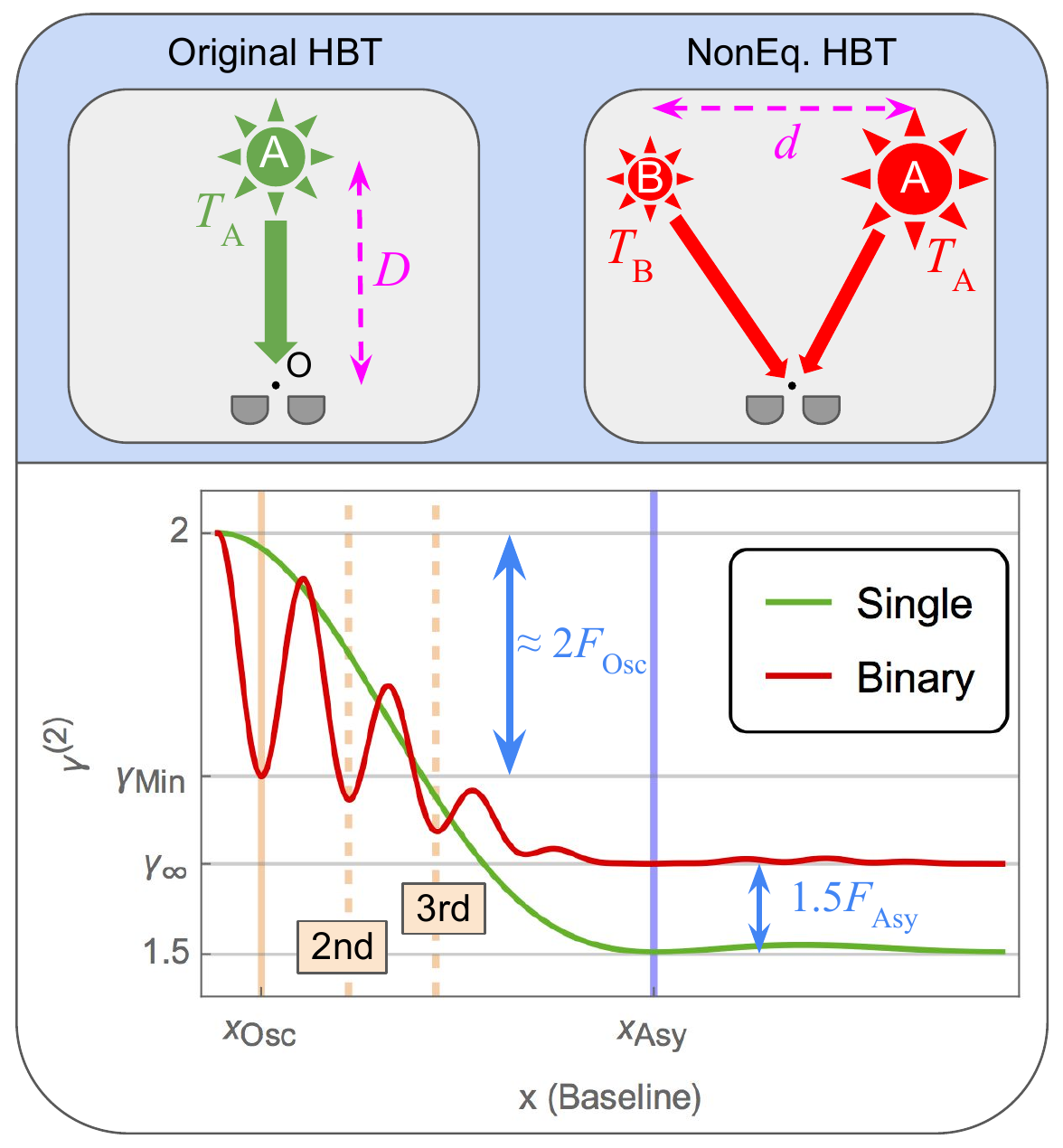}
\caption{\label{FigVision} Sketch highlighting the main features of the second-order coherence as a function of the separation betweeen two detectors $x=x_{1}-x_{2}$ (called `baseline') for a binary star in comparison to the single star case, corresponding to Eqs.(\ref{GammaBinary2nd}) and (\ref{GammaSingle}), respectively. The value at the minimum of the first oscillation $\gamma_{\rm Min}$ is given by Eq.(\ref{GammaMin}). The asymptotic value $\gamma_{\infty}$ is given by Eq.(\ref{GammaLimitLongDistance}). The variations of these two features with respect to the single case scenario are $F_{\rm Osc}$ and $F_{\rm Asy}$, defined in Section \ref{sec:Measurement}. The value $x_{\rm Asy}$ corresponds to the decay provided by the Bessel functions, given in Eq.(\ref{XBessel}), while $x_{\rm Osc}$ corresponds to the position of the first minimum, given by Eq.(\ref{XOsc}). As it is shown in the upper panel, the two objects in the binary scenario are assumed to have different temperatures $T_{\rm A,B}$ and sizes. Lastly, $D$ stands for the distance between the systems and the observation point O (around which the two detectors are located), while $d$ corresponds to the distance between the constituents of the binary system.}
\end{figure}

For the scenario of characterizing stars, this is a significant objective, since it is well-known that gravitationally-bounded system of stars are commonly found in the universe, for instance, as binary stars (system of two stars). Motivated by this but not restricted to it, the general scenario that we are interested in is formed by two extended spherical objects (A and B) of different radii and temperatures ($R_{\rm A,B}$ and $T_{\rm A,B}$, respectively) separated by a distance $d$ between them and by a distance $D$ with respect to the pair of detectors as shown in Fig.\ref{FigVision}.
The quantification of the coincidences is given by the second-order coherence associated to photo-counts on two detectors located at positions $\mathbf{x}_{1},\mathbf{x}_{2}$ \cite{gerry2005introductory}. The second-order coherence is defined in terms of the first- and second-order field correlations as $\gamma^{(2)}(\mathbf{x}_{1},\mathbf{x}_{2})\equiv G^{(2)}(\mathbf{x}_{1},\mathbf{x}_{2})/[G^{(1)}(\mathbf{x}_{1})G^{(1)}(\mathbf{x}_{2})]$, with $G^{(k)}(\mathbf{x}_{1},...,\mathbf{x}_{k})$ the $k^{th}$-order correlation of the field (see in the next Section and in App.\ref{App:GeneratingFunctional} for complete definitions). We employ the P-function representation for calculating the quantum field correlation as functional derivatives of a suitable generating functional. By summing the contributions of all the pairs of points over the cross-sections of the objects, we obtain a second-order coherence for extended sources at different temperatures. With respect to the single object scenario, the binary system presents a more complex second-order coherence function, as it is sketched in Fig.\ref{FigVision}. With respect to previous implementations of the HBT interferometry, we are incorporating the sizes but also the temperatures of the objects.

We show that the main differences between the single and binary cases are: 1- the appearance of oscillations, which in general shows its largest deviation from the single case in the first minimum ($\gamma_{\rm Min}^{(2)}$) which is found at a baseline $x=x_{\rm Osc}$; and 2- an asymptotic value ($\gamma_{\infty}^{(2)}$) larger than 1.5 found at baselines $x>x_{\rm Asy}$. Additionally, the feature $F_{\rm Osc}$ is defined as the difference between the first minimum and the single case value. Similarly, the $F_{\rm Asy}$ is the variation of the asymptotic value with respect to the single case. In some scenarios, minima with larger baselines will be useful too. These are located at baselines $(2m-1)x_{\rm Osc}$, with $m$ labelling them and limited by $(1+x_{\rm Asy}/x_{\rm Osc})/2$. The characterization of the binary system is given by the dependence of these quantities on the parameters of the system. In this work, we will show that for photons of frequency $\omega$ we have:
\begin{equation}
    \gamma_{\rm Min}^{(2)}=\gamma_{\rm Min}^{(2)}\left(\omega,s,T_{\rm A},T_{\rm B},d\right)~~,~~\gamma_{\infty}^{(2)}=\gamma_{\infty}^{(2)}\left(\omega,s,T_{\rm A},T_{\rm B}\right),
    \label{MinAsymptotic}
\end{equation}
\begin{equation}
    x_{\rm Osc}=x_{\rm Osc}\left(\omega,D,d\right)~~~,~~~x_{\rm Asy}=x_{\rm Asy}\left(\omega,D,R_{\rm A}\right),
    \label{OscBessel}
\end{equation}
where $s=(R_{\rm B}/R_{\rm A})^{2}$ is the surface ratio and $\omega=2\pi\nu$, being $\nu$ the frequency of collection. In this work we rigorously deduce the expression for Eqs. (\ref{MinAsymptotic}) and (\ref{OscBessel}). Furthermore, we show its implementation for the best possible characterization of a given binary system at different temperatures. We note that this technique can be applied to general scenarios, not limited to having the objects at different temperatures, but also to discern specific effects such as the relative orbital motion in astrophysical scenarios which is crucial for complete characterization of actual situations such as binary star systems.

This work is organized as follows: in Section \ref{sec:Theory} we fully develop our theoretical framework, deriving the general result for the second-order coherence and the expressions for the main features. We also analyze in this section the limiting cases, achieving insights and intuition on the underlying physical aspects. In Section \ref{sec:Measurement} we discuss strategical aspects for employing the main features in measurements. In Section \ref{sec:Orbit} we show how the orbital motion can be included in our calculations to some extent for some scenarios. In Section \ref{sec:Luhman16} we apply all our results to the case of binary stars, particularly analyzing the systems Luhman 16 and Spica. In Section \ref{sec:Lab} we discuss some of the key aspects for realizing experiments including nonequilibrium configurations. In Section \ref{sec:Conclusions} we summarize our findings and give some future insights on the applicability of our results. Finally, we devoted several appendices to show the intermediate steps of the calculations showed throughout the main text.

\section{\label{sec:Theory} General formalism for field correlations of binary systems}
%%%%%%%%%%%%%%%%%%%%%%%%%%%%%

In this section we develop the general theoretical framework for studying the electric field statistical properties. For this we investigate the electric field correlations associated to photo-measurements for studying the spatial coherence of the field generated by a specific sources configuration. In particular, we are interested in the correlation functions of a unique component of the electric field operator in such a way that $\hat{\mathbf{E}}=\hat{E}\check{\mathbf{e}}$.

As shown in App.\ref{App:GeneratingFunctional}, for studying spatial coherence, the correlation functions of interest can be written as:
\begin{eqnarray}
G^{(k)}(\mathbf{x}_{1},...,\mathbf{x}_{k},t)&\equiv&\left\langle \hat{E}^{(-)}(\mathbf{x}_{1},t)...\hat{E}^{(-)}(\mathbf{x}_{k},t)\right.\nonumber\\
&\times&\left.\hat{E}^{(+)}(\mathbf{x}_{1},t)...\hat{E}^{(+)}(\mathbf{x}_{k},t)\right\rangle.
\end{eqnarray}

These correlations can be derived from a generating functional $Z$ defined as:
\begin{eqnarray}
&&Z\left[a(\mathbf{x})\right]=\label{ZGeneratingEE}\\
&&=\left\langle\mathcal{T}:{\rm Exp}\left(\int d\mathbf{x}~a(\mathbf{x})\hat{E}^{(-)}(\mathbf{x},t)\hat{E}^{(+)}(\mathbf{x},t)\right):\right\rangle,\nonumber
\end{eqnarray}
where $:~:$ stands for the normal product of operators and $\mathcal{T}$ for the time-ordered product.

Then, the connection with the correlations is in terms of functional derivatives, having:
\begin{equation}
G^{(k)}(\mathbf{x}_{1},...,\mathbf{x}_{k},t)=\left.\frac{\delta^{k} Z}{\delta a(\mathbf{x}_{1})...\delta a(\mathbf{x}_{k})}\right|_{a=0}.
\label{QuantumCorrelationKDerivatives}
\end{equation}

For a specific scenario the generating functional allow us to obtain all the correlation functions connected to photo-counting detectors. The determination of the functional depends on the boundary conditions that enter through the electric field operator. We are interested in scenarios involving extended objects in the far-regime and describing the light coming from them. The electric field operator is given by a combination of plane waves% describing radiation coming from distant sources
:
\begin{equation}
\hat{\mathbf{E}}^{(+)}(\mathbf{x},\tau)=-i\sum_{\mathbf{p}\lambda}\left(\frac{2\pi\hbar\omega_{\mathbf{p}}}{V}\right)^{1/2}\hat{a}_{\mathbf{p}\lambda}e_{\mathbf{p}\lambda}~e^{i(\mathbf{p}\cdot\mathbf{x}-\omega_{\mathbf{p}}\tau)},
\label{ElectricFieldOperator}
\end{equation}
while $\hat{\mathbf{E}}^{(-)}(\mathbf{x},\tau)=[\hat{\mathbf{E}}^{(+)}(\mathbf{x},\tau)]^{\dag}$. The summation is over the modes of the electromagnetic (EM) field. This expansion agrees with the one for a quantized free field in a box (see Refs.\cite{Scully,Milonni}). The operator $\hat{a}_{\mathbf{p}\lambda}^{\dag}$ ($\hat{a}_{\mathbf{p}\lambda}$) corresponds to the creation (annihilation) operator of a photon of mode $\mathbf{p}\lambda$, being $\mathbf{p}$ the wave vector and $\lambda$ the polarization label and characterized by a frequency $\omega_{\mathbf{p}}=cp$, while $V$ corresponds to the volume for `box normalization'. The unit vector $e_{\mathbf{p}\lambda}$ accounts for the polarization of each plane wave, satisfying $\mathbf{p}\cdot e_{\mathbf{p}\lambda}=0$ while $\lambda=1,2$.% In the continuum limit, $\sum_{\mathbf{p}\lambda}\rightarrow[V/(2\pi)^{3}]\sum_{\lambda}\int d^{3}p$.

%%%%%%%%%%%%%%%%%%%%%%%%%%%%%
\subsection{\label{subsec:TwoSources}Nonequilibrium two-sources formalism}
%%%%%%%%%%%%%%%%%%%%%%%%%%%%%

First, we follow the approach of Ref.\cite{MandelReview} for the case of two point-sources (1 and 2) located at different points $P_{1,2}$ (characterized by positions $\mathbf{r}_{1,2}$) as in Young's interference experiment. Now, the specific form of the quantum state for two sources located at different positions must be introduced. Considering Glauber's representation, the state of one beam (coming from one of the sources) is represented in the basis $|\{v_{\beta_{1}}\}\rangle$ while the second one is analogously described in the basis $|\{v_{\beta_{2}}\}\rangle$, having $\beta=(\mathbf{p},\lambda)$. In addition, we assume the two point-sources to be independent each other. Then, the density operator of the combined field is given by:
\begin{eqnarray}
\hat{\rho}&=&\int\int d^{2}\{v_{\beta_{1}}\}d^{2}\{v_{\beta_{2}}\}\mathcal{P}_{1}\left(\{v_{\beta_{1}}\}\right)\mathcal{P}_{2}\left(\{v_{\beta_{2}}\}\right)\nonumber\\
&\times&\left|\{v_{\beta_{1}}\},\{v_{\beta_{2}}\}\right\rangle\left\langle\{v_{\beta_{2}}\},\{v_{\beta_{1}}\}\right|,
\label{StateGeneral}
\end{eqnarray}
where $\mathcal{P}_{i}(\{v_{\beta_{i}}\})$ corresponds to the Glauber's P-function of the $i-$th source (with $i=1,2$). This representation is still ambiguous since the `two-sources' states $|\{v_{\beta_{1}}\},\{v_{\beta_{2}}\}\rangle$ are not defined yet. In Ref.\cite{MandelReview} the authors declare how the field operator acts on this kind of states, following that:
\begin{eqnarray}
&&\hat{\mathbf{E}}^{(+)}(\mathbf{x},\tau)\left|\{v_{\beta_{1}}\},\{v_{\beta_{2}}\}\right\rangle=\label{TwoSourcesState}\\
&&=\left[\mathbf{E}^{(+)}_{1}(\mathbf{x},\tau)+\mathbf{E}^{(+)}_{2}(\mathbf{x},\tau)\right]\left|\{v_{\beta_{1}}\},\{v_{\beta_{2}}\}\right\rangle,\nonumber
\end{eqnarray}
having:
\begin{equation}
\mathbf{E}^{(+)}_{m}(\mathbf{x},\tau)=-i\sum_{\mathbf{p}\lambda}\sqrt{\frac{2\pi\hbar\omega_{\mathbf{p}}}{V}}v_{\mathbf{p}\lambda}e_{\mathbf{p}\lambda}~e^{i\left[\mathbf{p}\cdot\mathbf{r}_{m}-\omega_{\mathbf{p}}(\tau-t_{m})\right]},
\label{SourcesE}
\end{equation}
where $t_{m}=s_{m}/c$ is the time that it takes to a signal generated at $P_{m}$ to reach an observation point $P$ at $\mathbf{x}$. In this sense, $s_{m}=|\mathbf{x}-\mathbf{r}_{m}|$ corresponds to the distance between the source point $P_{m}$ and an observation point $P$. Thus, we can say that $\mathbf{E}_{1}+\mathbf{E}_{2}$ corresponds to the total electric field at the observation point $P$ due to the two point-sources.

\begin{figure}[!ht]
\includegraphics[width=\linewidth]{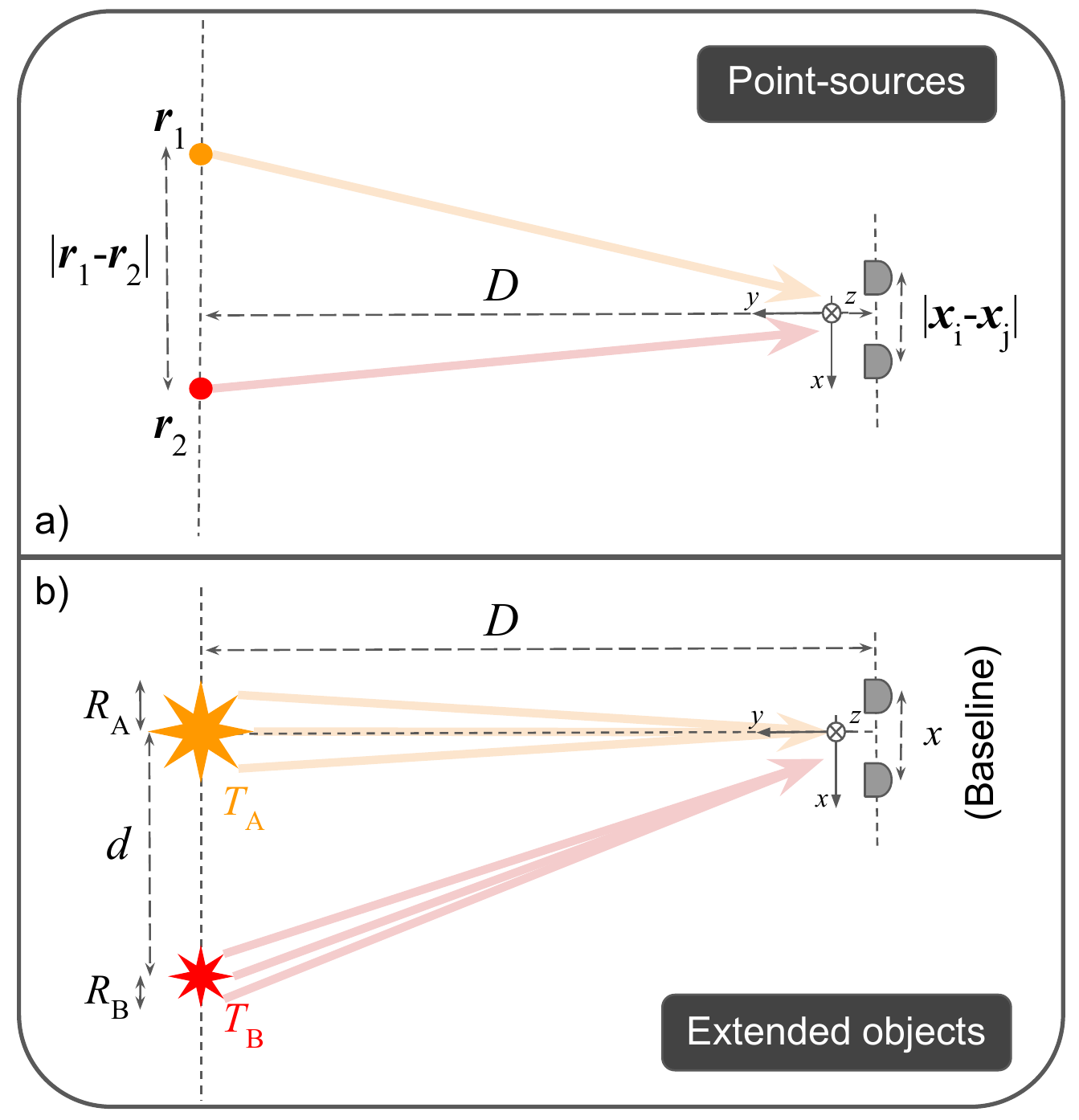}
\caption{\label{Fig1} Sketch of the scenarios. a) Two point sources at temperatures $T_{1}$ and $T_{2}$ and located at positions $\mathbf{r}_{1,2}$, respectively. The sketch shows the plane where the sources and the observation points (at positions $\mathbf{x}_{i,j}$, with $i=1,...,k$) are contained. Two sources are approximately located at a distance $D$ from the observation points and approximately aligned along the $x-$axis. The distance between the sources is given by $|\mathbf{r}_{1}-\mathbf{r}_{2}|$. In this scenario $D\gg|\mathbf{r}_{1}-\mathbf{r}_{2}|\gg|\mathbf{x}_{i}-\mathbf{x}_{j}|$, so each point-source radiates in a single wavevector. b) Two extended objects (EOs) at temperatures {\color{orange}{$T_{\rm A}$}} and {\color{red}{$T_{\rm B}$}}, for instance, stars or planets of radii $R_{\rm A,B}$. In contrast to the scenario described before, there are only two observation points (located at $\mathbf{x}_{1,2}$, respectively) and the distance between the centers of the EOs is $d$. In this scenario $D\gg d\gg x=|\mathbf{x}_{1}-\mathbf{x}_{2}|$. Given this, each EO radiates from several points but in approximately a single wavevector per point.}
\end{figure}

 We are interested in the quantum correlation functions for a scenario where two sources are separated from the observation points by a distance $D$ much larger than the distance between the observation points, as shown in Fig. \ref{Fig1}a. A frequency filtering allow us to consider just single modes arriving from each source. Each of the two modes have the same frequency $\omega$ but different wavevectors $\mathbf{p}_{i}=(\omega/c)\mathbf{n}_{i}$, being $\mathbf{n}_{i}$ the unit vectors associated to each direction connecting a point-source with the observation point $P$. For astrophysical scenarios, this is a reasonable assumption since the distance between the sources $|\mathbf{r}_{1}-\mathbf{r}_{2}|$ and the distance between all the observation points $|\mathbf{x}_{i}-\mathbf{x}_{j}|$ for all the pairs of observation points ($i,j=1,...,k$) satisfy $D\gg |\mathbf{r}_{1}-\mathbf{r}_{2}|\gg|\mathbf{x}_{i}-\mathbf{x}_{j}|$. %But this regime apply to many other scenarios within the Fraunhofer regime [REF]. 
Thus, the radiation coming from each source can be fairly approximated by a single mode with unit vectors $\mathbf{n}_{i}\equiv\mathbf{s}_{i}/s_{i}$ connecting each source to the observation point.
Furthermore, as $\mathbf{E}=E\mathbf{e}$, the electric fields of Eq.(\ref{SourcesE}) reads:
\begin{equation}
E^{(+)}_{m}(\mathbf{x},\tau)\approx-i\mathcal{E}_{0}v_{m}~e^{i\frac{\omega}{c}\left[\mathbf{n}_{m}\cdot\mathbf{r}_{m}-c(\tau-t_{m})\right]},
\label{ElectricSingleMode}
\end{equation}
with $\mathcal{E}_{0}=\sqrt{2\pi\hbar\omega/V}$.

Taking into account these considerations, the state in Eq.(\ref{StateGeneral}) can be given by the combination of states with just one mode per source. As we consider each source to be thermal, the P-functions associated to each one is:
\begin{equation}
    \mathcal{P}_{i}(v_{i})=\frac{1}{\pi\bar{n}_{i}}~{\rm Exp}\left[-\frac{|v_{i}|^{2}}{\bar{n}_{i}}\right],
    \label{P-FunctionThermal}
\end{equation}
where $\bar{n}_{i}=1/({\rm Exp}[\hbar\omega/(k_{\rm B}T_{i})]-1)$ is the mean photon number of the source $i$ at temperature $T_{i}$. Within the context of extended objects (EOs), let us remark that assuming a thermal state implies that each part of the objects are taken as black-body radiators. Strictly speaking, for astrophysical objects this is, of course, an approximation. However, for our purposes of studying the main features of photon correlations the approximation is sufficiently good.

Having all the previous considerations, we can write the generating functional for the scenario just described as:
\begin{eqnarray}
&&Z\left[a(\mathbf{x})\right]
=\\
&&=\frac{1}{\pi^{2}\bar{n}_{1}\bar{n}_{2}}\int\int e^{-\frac{|v_{1}|^{2}}{\bar{n}_{1}}-\frac{|v_{2}|^{2}}{\bar{n}_{2}}+\sum_{l,m=1}^{2}A_{ml}v_{l}^{*}v_{m}}d^{2}v_{1}d^{2}v_{2},\nonumber
\end{eqnarray}
with the coefficients given by:
\begin{equation}
A_{ml}\equiv\mathcal{E}_{0}^{2}%\frac{}{\epsilon_{0}}
\int d\mathbf{x}~e^{i\frac{\omega}{c}\Delta R_{ml}(\mathbf{x})}a(\mathbf{x}),
\end{equation}
where $\Delta R_{ij}(\mathbf{x})=R_{i}(\mathbf{x})-R_{j}(\mathbf{x})$ corresponds to the difference in the optical path traveled by the radiation coming from each source, so $R_{i}(\mathbf{x})=\mathbf{n}_{i}\cdot\mathbf{r}_{i}+ct_{i}$.

In App.\ref{App:GeneratingFunctionalTwoThermalSources} we show the complete calculation, obtaining:
\begin{equation}
Z\left[a(\mathbf{x})\right]
=\frac{1}{\left[\left(1-A_{11}\bar{n}_{1}\right)\left(1-A_{22}\bar{n}_{2}\right)-\bar{n}_{1}\bar{n}_{2}\left|A_{12}\right|^{2}\right]}.
\label{ZTwoSourcesFunctional}
\end{equation}

Employing Eq.(\ref{QuantumCorrelationKDerivatives}) we can obtain the correlation functions. In App.\ref{App:CorrelationsTwoThermalSources} we show the calculation for the first two correlation functions for a scenario in the far-regime, having:
\begin{equation}
G^{(1)}=%\bar{n}_{1}\left.\frac{\delta A_{11}}{\delta a(\mathbf{x}_{1})}\right|_{a=0}+\bar{n}_{2}\left.\frac{\delta A_{22}}{\delta a(\mathbf{x}_{1})}\right|_{a=0}=
\mathcal{E}_{0}^{2}%\frac{}{\epsilon_{0}}
\left(\bar{n}_{1}+\bar{n}_{2}\right),
\label{G1TwoThermalSources}
\end{equation}
\begin{eqnarray}
G^{(2)}(x_{1}-x_{2})&\approx&2\mathcal{E}_{0}^{4}%\frac{}{\epsilon_{0}^{2}}
\left(\bar{n}_{1}^{2}+\bar{n}_{2}^{2}+\bar{n}_{1}\bar{n}_{2}\right.\label{G2TwoThermalSources}\\
&\times&\left.\left[1+\cos\left(\frac{\omega [d_{1}-d_{2}]}{cD}[x_{1}-x_{2}]\right)\right]\right).\nonumber
\end{eqnarray}

Here, the first-order correlation is constant and independent of the sources' positions. This is expected since the two sources are incoherent, so the total intensity corresponds to the sum of the intensities of both sources. On the other hand, despite being incoherent sources, the intensity correlation of second-order provides further information as it depends on the distance between the sources ($d_{1}-d_{2}$), the distance to the sources' configuration with respect to the observation points ($D$) and the distance between detectors ($x_{1}-x_{2}$). Notice that for the case where the two sources are at the same position ($d_{1}=d_{2}$) we get $G^{(2)}_{d_{1}=d_{2}}(x_{1}-x_{2})=2(G^{(1)})^{2}$, which corresponds to the value for a thermal source of mean photon number $\bar{n}_{1}+\bar{n}_{2}$. The two sources held at different temperatures but at the same position are perceived as a single thermal source. The same value is obtained for the general case ($d_{1}\neq d_{2}$) when the two measurements are taken at the same point ($x_{1}=x_{2}$), having $G^{(2)}(0)=2(G^{(1)})^{2}$. The latter implies that measuring at a single point provides no further information about the sources' configuration than the one obtained from the first-order correlation $G^{(1)}$. However, different detection points ($x_{1}\neq x_{2}$) gives intensity correlations showing that temporal coincidence of photo-detections at different positions are affected by the difference in optical path traveled by the photons, finally depending on the distance between the sources.

%%%%%%%%%%%%%%%%%%%%%%%%%%%%%
\subsection{\label{subsec:SecondOrderCoherence}Second-order coherence for binary systems of extended objects at different temperatures}
%%%%%%%%%%%%%%%%%%%%%%%%%%%%%

As we mentioned in the Section \ref{sec:Intro}, a meaningful measure of the coincidences consists on the two point-sources second-order coherence %, defined as
$\gamma^{(2)}(\mathbf{x}_{1},\mathbf{x}_{2})%\equiv G^{(2)}(\mathbf{x}_{1},\mathbf{x}_{2})/[G^{(1)}(\mathbf{x}_{1})G^{(1)}(\mathbf{x}_{2})]
$. For the present case, we obtain:
\begin{eqnarray}
\gamma^{(2)}_{\rm 2S}(x_{1}-x_{2})&=&2\left(1+\frac{\mathcal{N}_{12}}{\left(1+\mathcal{N}_{12}\right)^{2}}\right.\label{Gamma22S}\\
&\times&\left.\left[\cos\left(\frac{\omega(d_{1}-d_{2})}{cD}(x_{1}-x_{2})\right)-1\right]\right),\nonumber
\end{eqnarray}
with $\mathcal{N}_{12}\equiv\bar{n}_{2}/\bar{n}_{1}$ the ratio between the photon average number of each point-source. The last expression is a generalization for two sources at different temperatures of the second-order coherence found in Ref.\cite{DifferentialHBT}. Notice that as a function of $x_{1}-x_{2}$, the second-order coherence is an oscillatory function, whose period is uniquely determined by the cosine's argument. The coincidence counts on a pair of detectors by photons coming from two sources depends on the distances between the sources ($d_{1}-d_{2}$), between the detectors ($x_{1}-x_{2}$) and from the detectors to the sources ($D$). This is a second-order interference effect. In contrary, two thermal sources present no interference pattern at first order, as it is shown from Eq.(\ref{G1TwoThermalSources}), so no amplitude interference can be exploit in order to get information about the sources' configuration. However, this kind of incoherence does not prevent higher order correlations to show a dependence with the source's relative position. Their intensities are correlated and show interference features of a pair of sources.

The case of extended of objects (EOs) is included by considering an array of point-sources emitting at a certain temperature from the objects' surfaces (see Fig.\ref{Fig1}.b). We assume the radiation field originated on the surfaces of the objects as a fair approximation. This connects to the high impenetrability of electromagnetic radiation on objects. Particularly, this approximation is fairly good for stars, but it stands as an assumption clearly beyond that.

Given a configuration of EOs, each pair of points taken from them presents a second-order coherence given by Eq.(\ref{Gamma22S}). The total second-order coherence for the configuration is obtained as a sum of the second-order coherences of all the pairs of points emitting light that reaches the observation points.

For the case of a binary system of EOs we have to integrate over the surfaces of the constituents $S_{\rm A,B}$:
\begin{eqnarray}
\gamma_{\rm Binary}^{(2)}(\mathbf{x}_{1},\mathbf{x}_{2})&=&\frac{1}{(S_{\rm A}+S_{\rm B})^{2}}\int_{S_{\rm A}\cup S_{\rm B}}dS_{\mathbf{r}_{1}}\int_{S_{\rm A}\cup S_{\rm B}}dS_{\mathbf{r}_{2}}\nonumber\\
&\times&\gamma^{(2)}(\mathbf{x}_{1},\mathbf{x}_{2},\mathbf{r}_{1},\mathbf{r}_{2}),
\label{Binary2ndCoherence}
\end{eqnarray}
where we have written the explicit dependence of the second-order coherence on the sources' positions to avoid confusions.

In App.\ref{App:SecondOrderCoherenceBinary} we show the full calculation of the second-order coherence for a binary system of two spherical EOs. A crucial approximation is that the surfaces $S_{\rm A,B}$ are taken as discs of radii $R_{\rm A,B}$. Since $D\gg |\mathbf{r}_{1}-\mathbf{r}_{2}|\gg|\mathbf{x}_{i}-\mathbf{x}_{j}|$, the curvature of the objects with respect to the observers is negligible so the integration can be taken over the flat cross section. Thus, the surfaces are taken as $S_{\rm A,B}\approx\pi R_{\rm A,B}^{2}$. Given that the separation between the centers of the constituent is $d$, the second-order coherence for the binary system results:
\begin{eqnarray}
\gamma_{\rm Binary}^{(2)}(x_{1}-x_{2})&=&\gamma_{\infty}^{(2)}+\frac{1}{(1+s)^{2}}\left[\gamma_{\rm AA}^{(2)}(x_{1}-x_{2})\right.\nonumber\\
&+&\left.s^{2}\gamma_{\rm BB}^{(2)}(x_{1}-x_{2})
+\frac{8\mathcal{N}s}{(1+\mathcal{N})^{2}}\right.\label{GammaBinary2nd}\\
&\times&\left.\cos\left(\frac{\omega d}{cD}[x_{1}-x_{2}]\right)\gamma_{\rm AB}^{(2)}(x_{1}-x_{2})\right],\nonumber
\end{eqnarray}
with each contribution given by:
\begin{equation}
\gamma_{\infty}^{(2)}=\frac{3}{2}+\frac{s}{(1+s)^{2}}\frac{(1-\mathcal{N})^{2}}{(1+\mathcal{N})^{2}},
\label{GammaLD}
\end{equation}
\begin{eqnarray}
\gamma_{\rm ij}^{(2)}(x_{1}-x_{2})&=&\frac{2}{R_{i}R_{j}}\left[\frac{cD}{\omega(x_{1}-x_{2})}\right]^{2}\label{GammaIJ}\\
&\times&J_{1}\left(\frac{\omega R_{i}}{cD}[x_{1}-x_{2}]\right)J_{1}\left(\frac{\omega R_{j}}{cD}[x_{1}-x_{2}]\right),\nonumber
\end{eqnarray}
where $s=S_{\rm B}/S_{\rm A}=(R_{\rm B}/R_{\rm A})^{2}$ is the surface ratio and $\mathcal{N}=\bar{n}_{\rm B}/\bar{n}_{\rm A}$ the ratio between the mean photon numbers at the temperature of each object. Finally, $x_{1}-x_{2}=x$ results the distance between the detectors, known as the \emph{baseline}.

%%%%%%%%%%%%%%%%%%%%%%%%%%%%%
\subsection{\label{subsec:GeneralFeatures}Characterization of binary systems: Oscillations and asymptotic value}
%%%%%%%%%%%%%%%%%%%%%%%%%%%%%

Having obtained a general expression for the second-order coherence for arbitrary baselines, we proceed to characterize binary systems. We focus on the main differences of the second-order coherence with respect to the single case, as we anticipated in Section \ref{sec:Intro}. The main features are sketched in the lower panel of Fig.\ref{FigVision}. While the single object scenario is only characterized by a decay from 2 to 1.5 for a baseline $x_{\rm Asy}$ with no direct dependence on the temperature of the source, the binary system scenario presents features that depend on the parameters of the constituents $\{R_{\rm A,B},T_{\rm A,B}\}$. This is the result of interference effects affecting the photons distributions and, therefore, the coincidence counts on the detectors. One feature corresponds to the oscillations of frequency $2x_{\rm Osc}$. The first minimum occurs for a baseline $x=x_{\rm Osc}$, whose value of the second-order coherence is $\gamma_{\rm Min}^{(2)}\equiv\gamma^{(2)}(x_{\rm Osc})$. This defines an amplitude of the oscillation for the shortest meaningful baseline. % As we show below, this amplitude depends on the surfaces ratio $s%=(R_{\rm B}/R_{\rm A})^{2}
%$, both temperatures $T_{\rm A,B}$ and also on the distance between the objects $d$. The amplitude maximizes when is  $\gamma_{\rm Min}^{(2)}=1.5$, which happens for identical sources ($T_{\rm A}=T_{\rm B}$ and $R_{\rm A}=R_{\rm B}$ simultaneously).
A second crucial feature corresponds to the long-baseline or asymptotic value $\gamma_{\infty}^{(2)}$ at which the function decays for $x>x_{\rm Asy}$. Instead of decaying to 1.5, as in the single object case, the function can take different values for some systems.

The features and limiting expressions can be summarized as follows:
\\

1 - For $x_{1}-x_{2}=0$ we have $\gamma_{\rm Binary}^{(2)}(0)=2$ regardless on the system's parameters. This is expected since the involved sources are thermal. The second-order coherence for equal time and same position of the detectors must give the well-known value for thermal light.

2 - Notice that in the limit of small radius of the companion object $R_{\rm B}\rightarrow 0$, we have:
\begin{equation}
\gamma_{\rm Binary}^{(2)}(x_{1}-x_{2})\rightarrow\gamma_{\rm Single}^{(2)}(x_{1}-x_{2}),
\end{equation}
obtaining the single star result, given by:
\begin{equation}
\gamma_{\rm Single}^{(2)}(x_{1}-x_{2})=\frac{3}{2}+2\left[\frac{cD ~J_{1}\left(\frac{\omega R_{\rm A}}{cD}[x_{1}-x_{2}]\right)}{\omega R_{\rm A}(x_{1}-x_{2})}\right]^{2}.
\label{GammaSingle}
\end{equation}

3 - For the limit of large distances between the detectors (large values of $x_{1}-x_{2}$), we have the asymptotic value:
\begin{equation}
\gamma_{\rm Binary}^{(2)}(x_{1}-x_{2})\rightarrow\gamma_{\infty}^{(2)},
\label{GammaLimitLongDistance}
\end{equation}
with $3/2<\gamma_{\infty}^{(2)}<2$ and which only depends on the radii ($R_{\rm A,B}$) and the temperatures ($T_{\rm A,B}$) of each object, and on the frequency $\omega$. The value $\gamma_{\infty}^{(2)}=3/2$ is achieved for the equilibrium case ($\mathcal{N}\equiv 1$). A binary system at thermal equilibrium presents the same asymptotic value as a single object. The dependence on $s$ and $T_{\rm A,B}$ is such that similar radii but nonequilibrium gives the opportunity to obtain information about the binary system from this quantity.

Furthermore, let us remark that there is no dependence on the observation points ($x_{1,2}$), the distance to the binary system ($D$) and the distance between the components ($d$). The decay to $\gamma_{\infty}^{(2)}$ occurs for $x=x_{\rm Asy}$, with:
\begin{equation}
    x_{\rm Asy}=\frac{u_{1}cD}{\omega R_{\rm A}}.
    \label{XBessel}
\end{equation}
corresponding to the first zero of the Bessel function ($u_{1}\approx 3,83...$) and containing the largest radius $R_{\rm A}$. We will refer to $x_{\rm Asy}$ as the decay baseline.

4 - The oscillating behavior of the two point-sources case [see below Eq.(\ref{Gamma22S})] is inherited by the pair of EOs as oscillations limited by the decay baseline. The period of the oscillations is given by the oscillation baseline:
\begin{equation}
    x_{\rm Osc}=\frac{\pi cD}{\omega d},
    \label{XOsc}
\end{equation}
having $x_{\rm Osc}<x_{\rm Asy}$ since $R_{\rm A}<d$. Remarkably, this is the only feature that depends on $d$. As it was mentioned in Section \ref{sec:Intro}, in some situations the minima with larger baselines are useful. Their baselines are given by $(2m-1)x_{\rm Osc}$. As long as $m\leq(1+x_{\rm Asy}/x_{\rm Osc})/2$, a minimum takes place.

For the first of these oscillations we have $\gamma_{\rm Min}^{(2)}\equiv\gamma^{(2)}_{\rm Binary}(x_{\rm Osc})$, allowing for a maximal value of the ratio $\gamma^{(2)}_{\rm Binary}(x)/\gamma^{(2)}_{\rm Single}(x)$. In addition, having $R_{i}/d\ll 1$, we can show:
\begin{equation}
\gamma_{\rm ij}^{(2)}(x_{\rm Osc})=\frac{2}{R_{i}R_{j}}\left[\frac{d}{\pi}\right]^{2}J_{1}\left(\frac{\pi R_{i}}{d}\right)J_{1}\left(\frac{\pi R_{j}}{d}\right)\approx\frac{1}{2},
\label{GammaIJ}
\end{equation}
which gives:
\begin{eqnarray}
    \gamma_{\rm Min}^{(2)}&\approx&2-8\frac{s}{(1+s)^{2}}\frac{\mathcal{N}}{(1+\mathcal{N})^{2}}.
    \label{GammaMin}
\end{eqnarray}

Notice that the minimum possible value is $\gamma_{\rm Min}^{(2)}=3/2$, implying a maximal oscillation amplitude. This value is reached for the case of identical EOs, such that $s=1$ (same sizes) and $\mathcal{N}=1$ (thermal equilibrium). If we just impose the thermal equilibrium condition, we get oscillations provided that $s\neq 0$ but without maximal amplitude. The same happens in general, for a scenario of objects with different sizes and temperatures. From just the oscillation amplitude these two scenarios cannot be distinguished.

Also, notice that for a scenario where $x_{\rm Asy}\gg x_{\rm Osc}$, $\gamma^{(2)}_{\rm Binary}([2m-1]x_{\rm Osc})\approx\gamma_{\rm Min}^{(2)}$ for $m>1$ but not for all of them. This allows to employ the larger baselines $(2m-1)x_{\rm Osc}$ for eventually measure the a maximal oscillation amplitude.

5 - A limit of point-sources is obtained by setting $R_{\rm A}=R_{\rm B}\equiv R$ and then taking $R\rightarrow 0$, so:
\begin{equation}
\gamma_{\rm Binary}^{(2)}(x_{1}-x_{2})\rightarrow %2+\frac{r}{\left(1+r\right)^{2}}\left[\cos\left(\frac{\omega[d_{1}-d_{2}]}{cD}[x_{1}-x_{2}]\right)-1\right]\nonumber\\
%&=&
1+\frac{\gamma_{\rm 2S}^{(2)}(x_{1}-x_{2})}{2}%\nonumber\\
%&=&\frac{\gamma_{\rm Th}^{(2)}+\gamma_{\rm 2S}^{(2)}(x_{1}-x_{2})}{2}
,
\label{Gamma2BinaryEqualRadiusLimit}
\end{equation}
replacing $d_{1}-d_{2}$ by $d$ in Eq.(\ref{Gamma22S}).

This simplified expression is effective for systems of objects with similar sizes ($R_{\rm A}\approx R_{\rm B}$). The last relation results to be a connection between a limiting case for the binary system and the two point-sources scenario. Nevertheless, the binary system in the limit $R_{\rm A}=R_{\rm B}\equiv R\rightarrow 0$ remains different from the two point-sources systems although they share some aspects on the behavior of its second-order coherences.
\\

Notice that in points 3 and 4 we are showing explicitly the dependencies as shown in Eqs.(\ref{MinAsymptotic}) and (\ref{OscBessel}).

All in all, the second-order coherence is a bounded curve that starts at $\gamma_{\rm Binary}^{(2)}(0)=2$ and decay to $\gamma_{\infty}^{(2)}$. Depending on the predominance of the last term in Eq.(\ref{GammaBinary2nd}), oscillations can be presented in the second-order coherence function. For the limiting case, $d=0$, we get the result for the second-order coherence when the two objects have their centers in the same point. The photons coming from both sources travel approximately the same distance to each observation points. For this case, there are no oscillations and the second-order coherence is a monotonous decreasing function. For $d\neq 0$, oscillations arise and they can have an impact in the general form of the second-order coherence. As the total system presents two EOs, the pattern appears to be fringes with an envelope as in diffraction phenomena due to the finite size of the objects.

In this Section, we have described the physical aspects of the second-order coherence for a binary system. We showed as main features the oscillation amplitude of the first minimum and the asymptotic value. In the end of point 4, we have briefly discussed a limitation of measuring just the oscillation amplitude. For some scenarios, it is not enough for an appropriate characterization. Nevertheless, the issue can be solved by additionally verifying if $\gamma_{\infty}^{(2)}=3/2$ holds, which is true for every $s$ in an equilibrium scenario ($\mathcal{N}=1$). In this sense, this is a remarkable example where complementing both features might be useful. We have not analyzed yet how to choose the best strategy for measurements and describe the interplay between them. This is the aim of the next section.

%%%%%%%%%%%%%%%%%%%%%%%%%%%%%
\section{\label{sec:Measurement}Oscillation amplitude vs. asymptotic value: Competition or complementarity}
%%%%%%%%%%%%%%%%%%%%%%%%%%%%%

In previous sections we have given the second-order coherence for a binary system. We have shown that the two main features are the oscillation amplitude associated to the first minimum $\gamma_{\rm Min}^{(2)}$ and the asymptotic value $\gamma_{\infty}^{(2)}$ [Eqs.(\ref{GammaMin}) and (\ref{GammaLimitLongDistance}), respectively]. Both depend on the frequency, the radii and the temperatures of the objects. At the same time, we have that each feature occurs at different scales, given by the oscillation baseline $x_{\rm Osc}$ and the decay baseline $x_{\rm Asy}$ [Eqs.(\ref{XOsc}) and (\ref{XBessel}), respectively]. Only the former depends on the separation between the constituent objects, $d$. In principle, each scale can be adjusted by the frequency of observation. In many scenarios, it is common to have prior information of the system obtained from independent complementary methods, such as spectral and visual observations. For the purpose of the present analysis, we will consider $\{R_{\rm A},T_{\rm A}\}$ as known parameters for the rest of this section. Then, a natural concern is to analyze the competition but also the complementarity of the two features over the variety of possible scenarios. In this sense, it is imperative to point out the crucial aspects to consider for perfoming measurements with existing telescopes. Furthermore, we show how to compare the features for deciding which one is optimal to measure or to point out when do the features complement each other for achieving the best estimation of the properties of a system.

For addressing these matters, two aspects are crucial:

\textbf{(\textit{i}).} The comparison between the variations of each quantity with respect to the single star case. For the case of the oscillation amplitude, the variation is given by $F_{\rm Osc}=1-\gamma_{\rm Min}^{(2)}/\gamma_{\rm Single}^{(2)}(x_{\rm Osc})$, while for the asymptotic value the variation is $F_{\rm Asy}=2\gamma_{\infty}^{(2)}/3-1$.

\textbf{(\textit{ii}).} The largest variation has to be appreciable.

%Point \textit{i} implies the direct comparison between the variations. This gives which feature is better for measurements. However, this comparison does not guarantee that the largest variation is observable. This is precisely what point \textit{ii} addresses. 
For point \textbf{(\textit{ii})}, we take the criteria that a variation is appreciable if it is larger than $1\%$. For both variations $\{F_{\rm Osc},F_{\rm Asy}\}$ we obtain the same bounds on the surface ratio $s$:
\begin{equation}
    \frac{1}{10}<\frac{R_{\rm B}}{R_{\rm A}}<10,%10^{-2}<s<10^{2},
    \label{SizeRequirement}
\end{equation}
with no dependence on rest of the parameters. For observing $F_{\rm Osc}$ or $F_{\rm Asy}$, the size of the two sources must be similar up to one order of magnitude. Given that $\gamma_{\rm Min}^{(2)}$ and $\gamma_{\infty}^{(2)}$ depend on $s/(1+s)^2$ [Eqs. (\ref{GammaMin}) and (\ref{GammaLD}), respectively], the maximum value is reached for $s=1$. Given a pair of temperatures, sources of equal sizes maximize both variations.

While point \textbf{(\textit{ii})} gives bounds on $s$, the direct comparison of point \textbf{(\textit{i})} gives restrictions on the temperatures ($T_{\rm A,B}$) and the frequency $\omega$. Additionally taking $\omega$ as known, from %$1-\gamma_{\rm Min}^{(2)}/\gamma_{\rm Single}^{(2)}(x_{\rm Osc})=2\gamma_{\infty}^{(2)}/3-1$
$F_{\rm Osc}=F_{\rm Asy}$, we obtain that $T_{\rm B}=T_{\pm}=\hbar\omega/[k_{\rm B}{\rm ln}(1+1/[(4\pm\sqrt{15})\bar{n}_{\rm A}])]$. Thus, the conditions on $T_{\rm B}$ are:
\begin{equation}
    T_{-}<T_{\rm B}<T_{+}\implies F_{\rm Osc}>F_{\rm Asy
    }%{\text{Oscillations variation larger}}
    ,\label{OscillationRegion}%\\
%    T_{-}<T_{\rm B}<T_{+}\implies{\text{Asymptotic variation larger}}\nonumber,
\end{equation}
while the opposite is true for $T_{\rm B}$ outside that region.

\begin{figure}[!ht]
\includegraphics[width=\linewidth]{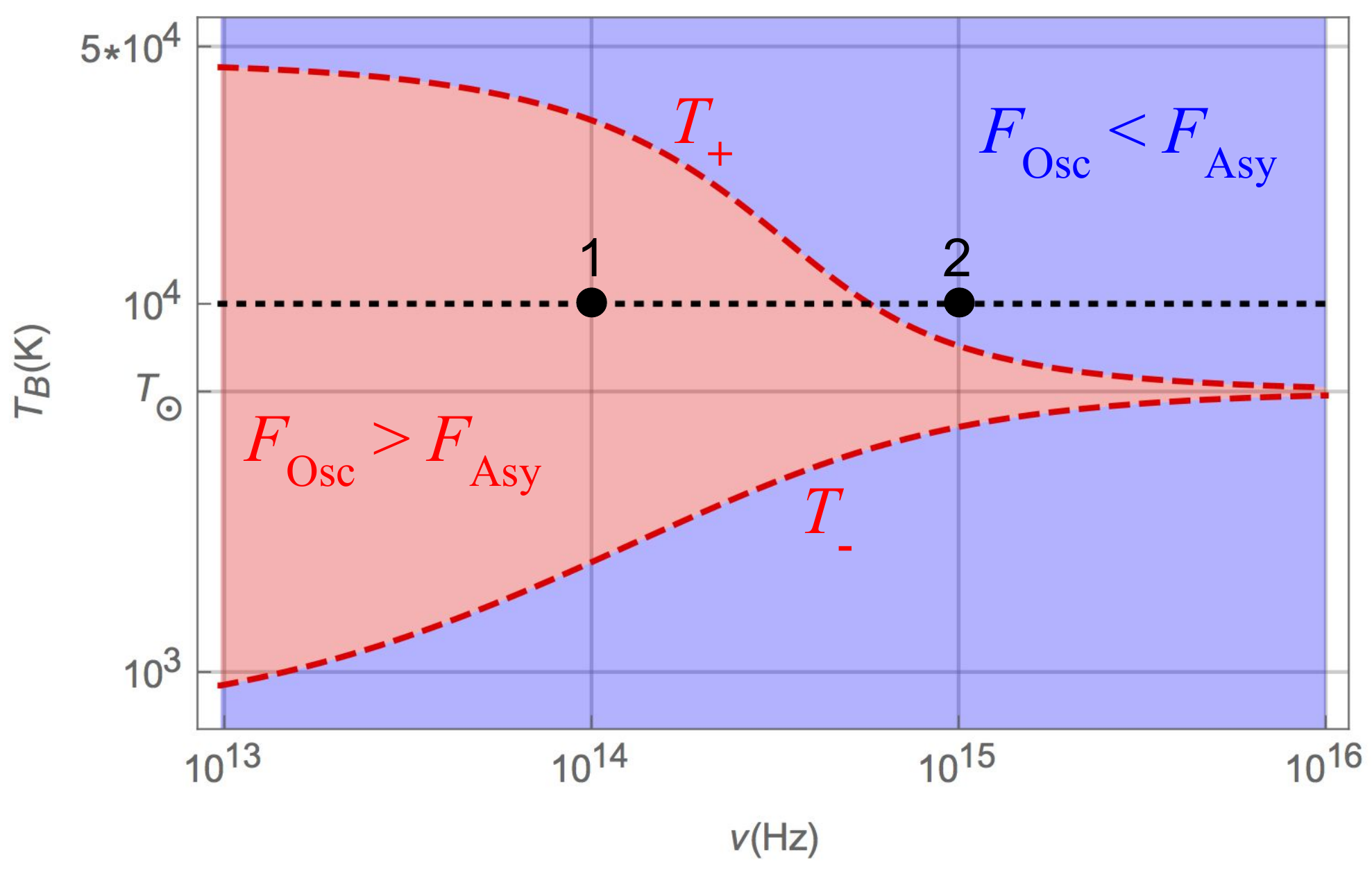}
\caption{\label{FigTBVsFreqDecision} Decision plot for a binary system with $T_{\rm A}=T_{\odot}=5778{\rm K}$. The upper (lower) red solid curve corresponds to $T_{+}(T_{-})$. The red shaded region corresponds to the inequalities given in Eq.(\ref{OscillationRegion}), while the blue corresponds to the opposite condition. The black dotted line corresponds to $T_{\rm B}=10^{4}{\rm K}$, while the value $T_{\rm B}=T_{\odot}$ corresponds to the equilibrium temperature. Black point 1 (2) corresponds to $\nu=10^{14}(10^{15}){\rm Hz}$ and $T_{\rm B}=10^{4}{\rm K}$. While for point 1 it happens that $F_{\rm Osc}>F_{\rm Asy}$, for point 2 the opposite is true.}
\end{figure}

Figure \ref{FigTBVsFreqDecision} corresponds to a plot with axis $\{\nu,T_{\rm B}\}$ (where $\omega=2\pi\nu$) that comprises the decision of which feature dominates. On the plot we consider a system where $T_{\rm A}=T_{\odot}=5778{\rm K}$.
The red shaded region corresponds to the double inequality given in Eq.(\ref{OscillationRegion}), bounded by the curves $T_{\pm}$. In the blue shaded region, $F_{\rm Asy}>F_{\rm Osc}$. Notice that $F_{\rm Osc}$ is always larger when the temperatures of the two constituents are similar ($T_{\rm B}\approx T_{\odot}$ for the case of the figure). This is expected since the oscillations are interference effects connected to the coherence and the bunching of photons. These are enhanced due to the similarity of the two sources. On the other hand, $F_{\rm Asy}$ becomes important for greater thermal imbalances.

\begin{figure}[!ht]
\includegraphics[width=\linewidth]{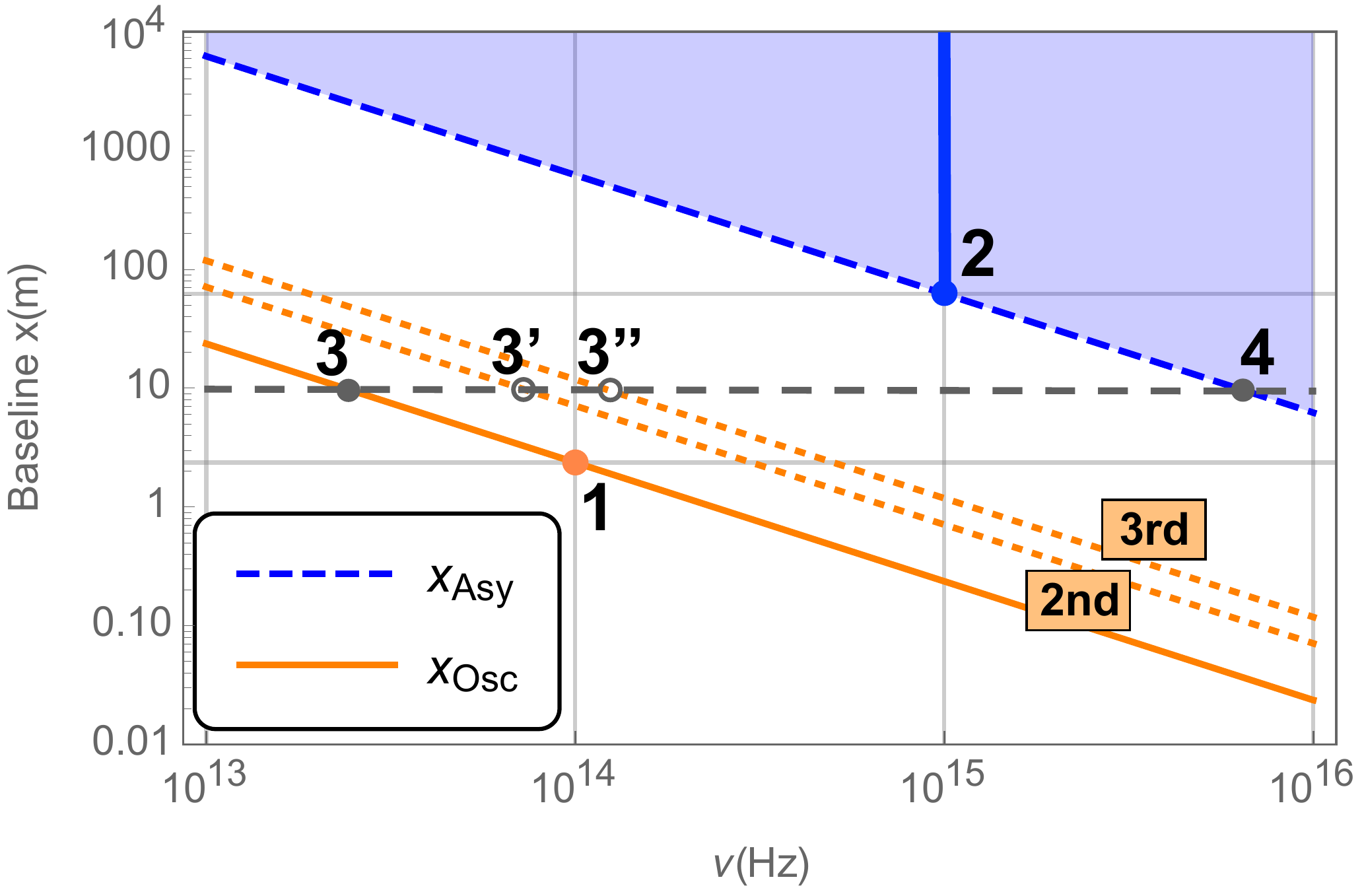}
\caption{\label{FigXVsFreq} Blue dashed (orange) curve corresponds to the decay (oscillation) baselines $x_{\rm Asy}$ ($x_{\rm Osc}$) according to Eq.(\ref{XBessel}) [Eq.(\ref{XOsc})] as a function of the frequency for a binary star located at $D=25{\rm ly}$, with the radius $R_{\rm A}=R_{\odot}$ and a distance between the constituents given by $d=1{\rm AU}$. The blue shaded region corresponds to the region $x>x_{\rm Asy}$. The orange dotted lines corresponds to the second and third minimum of oscillation, respectively. Point 1 (2) corresponds to $\nu=10^{14}(10^{15}){\rm Hz}$ shown in Fig.\ref{FigTBVsFreqDecision}. The blue solid line corresponds to the possible baselines at $\nu=10^{15}{\rm Hz}$. While for point 1 the baseline is in the order of meters, for point 2 is at least one order of magnitude larger. Dark gray dashed horizontal line corresponds to a fix baseline of 10m. All the minima and the asymptotic value can be measured with a baseline of 10m by changing the frequency of detection. Point 3, 3' and 3'' corresponds to the minima obtained at different frequencies, while point 4 corresponds to measuring the asymptotic value.}
\end{figure}

Note that the frequency of measurement determines which variation $\{F_{\rm Osc},F_{\rm Asy}\}$ is greater. For instance, consider the second object at a temperature $T_{\rm B}=10^{4}{\rm K}$ (black dotted line in Fig.\ref{FigTBVsFreqDecision}). Whilst, for frequency  $10^{14}{\rm Hz}$ (given by point 1) $F_{\rm Osc}>F_{\rm Asy}$. On the contrary, for frequency $10^{15}{\rm Hz}$ (given by point 2) the opposite is true ($F_{\rm Osc}<F_{\rm Asy}$). In each case, the strategy for the measurements is different. While for the former we require baselines of the order of the oscillation baseline $x_{\rm Osc}$, for the latter the baseline has to be larger than the decay baseline $x_{\rm Asy}$. Figure \ref{FigXVsFreq} shows the decay and oscillation baselines for $D=25{\rm ly}$, $R_{\rm A}=R_{\odot}$ and $d=1{\rm AU}$.
We can observe that for $10^{14}$Hz (point 1 in Fig.\ref{FigTBVsFreqDecision}) the baselines have to be of the order of a few meters for measuring $\gamma_{\rm Min}^{(2)}$. Moreover, for a given frequency $\nu$ when $x_{\rm Osc}(\nu)\ll x_{\rm Asy}(\nu)$ is true, then $\gamma^{(2)}_{\rm Binary}([2m-1]x_{\rm Osc})\approx\gamma_{\rm Min}^{(2)}$, for some $m>1$. In Fig.\ref{FigXVsFreq} the orange dotted curves corresponds to $m=2,3$. We can use the values of these larger baselines at a given frequency for measuring the oscillation amplitude. For point 2 the baselines has to be at least of $50{\rm m}$ for measuring $\gamma_{\infty}^{(2)}$. Any larger baseline at this frequency also works for measuring the asymptotic value, as denoted by the vertical solid blue line in Fig.\ref{FigXVsFreq}. Also, notice that for a baseline of 10m (dashed black horizontal line) it is possible to measure alternatively both features by changing the observation frequency. Maximal amplitudes will be obtained at Points 3, 3' and 3'', for different observation frequencies. For the asymptotic value, a larger frequency is required, corresponding to Point 4.

\begin{figure}[!ht]
\includegraphics[width=\linewidth]{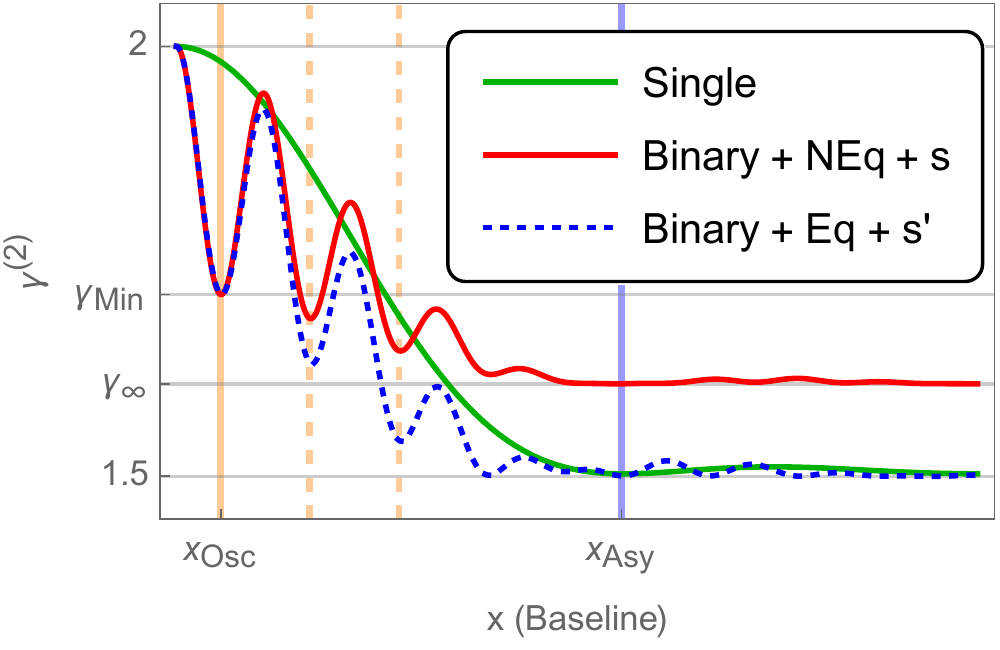}
\caption{\label{FigExampleDistinguish} Sketch highlighting how the main features of the second-order coherence change as a function of the baseline (separation between the detectors) $x=x_{1}-x_{2}$ for a scenario at equilibrium ($T_{\rm A}\simeq T_{\rm B}$) and with surface ratio $s'$ (blue dotted curve) and a nonequilibrium ($T_{\rm A}\neq T_{\rm B}$) with a surface ratio $s\neq s'$. The parameteres were chosen such that both scenarios presents the same oscillation amplitude for the first minimum. Having no prior about the temperatures, a distinction between the two scenarios results from a complementation between the measurements of both features (oscillation amplitude and asymptotic value). As before, the green one correspond to the single case. The vertical lines corresponds to the baseline values associated to $x_{\rm Asy}$ (blue solid), $x_{\rm Osc}$ (orange solid) and the next two minimum (orange dashed).}
\end{figure}

In the above arguments, we considered the scenario wherein the temperature of the first object is known. But, of course, another possible scenario is when no prior information is available about both the objects involved. In this case, it is useful to distinguish between an equilibrium scenario ($T_{\rm A}\simeq T_{\rm B}$) from a nonequilibrium one ($T_{\rm A}\neq T_{\rm A}$). If we are facing an equilibrium scenario without prior information, the complementarity of the features can be exploited as the best strategy. To illustrate this, Fig.\ref{FigExampleDistinguish} shows two scenarios, one at equilibrium and with a surface ratio $s'$ (blue dotted curve) and a second one in nonequilibrium and surface ratio $s\neq s'$ (red solid curve), both characerized by the same oscillation amplitude at the first minimum.
First, notice that at equilibrium, $F_{\rm Asy}=0$ irrespective of the surface ratio $s'$. Measuring $\gamma_{\infty}^{(2)}=3/2$ reveals that the scenario is an equilibrium scenario, but it will give no further information about any parameter of the system. Alternatively, if we only proceed to measure $\gamma_{\rm Min}^{(2)}$ without prior information about the temperatures, then the two scenarios shown in Fig.\ref{FigExampleDistinguish} are compatible with the measurements. A solution for a given $s$ and $T_{\rm B}=T_{\rm A}$ might be compatible with the measured value $\gamma_{\rm Min}^{(2)}$. In this case, both temperatures will remain unknown since for $T_{\rm B}\simeq T_{\rm A}$ it happens that $\gamma_{\rm Min}^{(2)}$ is just a function of $s'$. A second possible scenario compatible with the same measured value $\gamma_{\rm Min}^{(2)}$ could be found under the condition $T_{\rm B}\neq T_{\rm A}$ by choosing a suitable value of $s\neq s'$. Thus, without prior information about the temperatures of the objects, a complementary measurement of $\gamma_{\infty}^{(2)}$ will lead to a distinction between the two. In this sense, both features complement each other to uniquely distinguish a scenario.

Until this point, we have analyzed the correlations of the photons coming from two sources at a fixed distance $d$. But it is common in binary systems, particularly in astrophysics, that one of the objects is orbiting the other one due to its gravitational interaction. In this sense, it is crucial to include the orbital motion, at least in an approximate way, to consider its impact on the photon correlations. In the next section, we show how this can be done in a simple way for the case where a circular orbit is contained on the plane shown in Fig.\ref{Fig1}.

%%%%%%%%%%%%%%%%%%%%%%%%%%%%%
\section{\label{sec:Orbit}Orbital motion inclusion}
%%%%%%%%%%%%%%%%%%%%%%%%%%%%%

\begin{figure*}[!ht]
\includegraphics[width=0.7\linewidth]{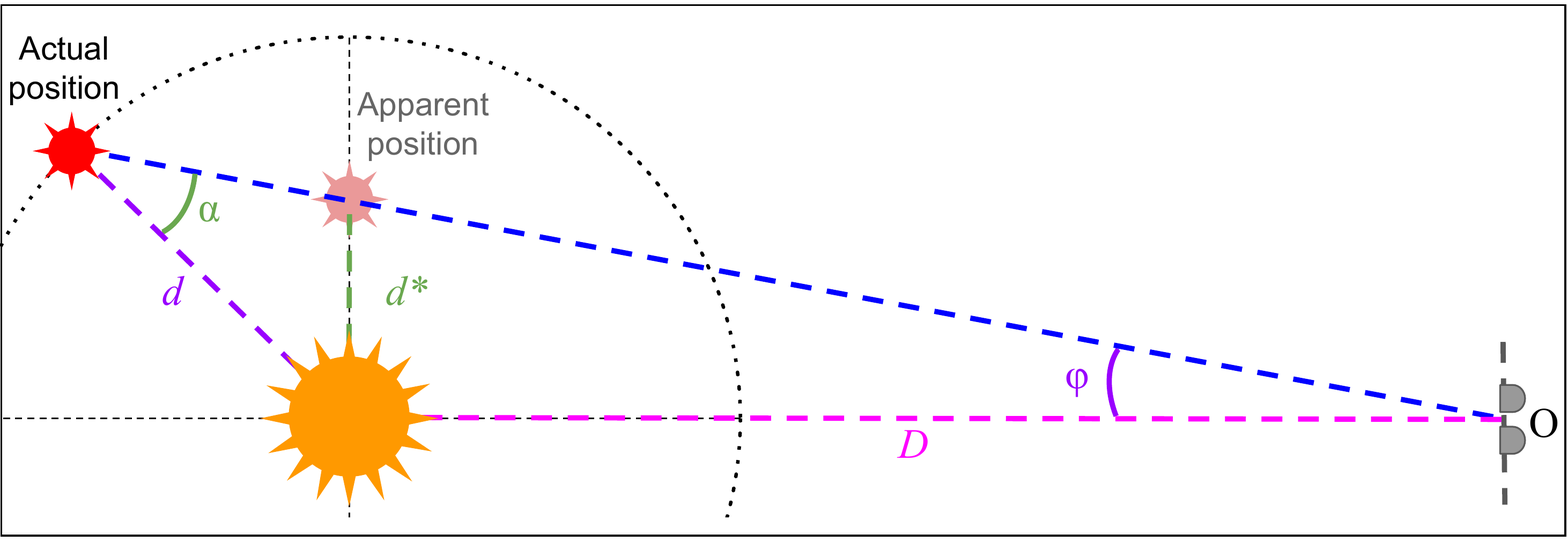}
\caption{Sketch for the definition of the apparent distance $d^{*}$ between the two EOs with respect to the observer $O$ given that $D\gg d$. The angle $\alpha$ corresponds to the phase angle, while the $\varphi$ is the angular separation between the EOs.}
\label{FigApparentd}
\end{figure*}

As we mentioned before, the effect of the orbital motion should be included for a complete analysis of, for instance, stellar systems. In these scenarios, the orbital motion is assumed as circular to a good approximation. A replacement of the constituents' actual distance $d$ by its apparent distance $d^{*}$ is appropriate since $D\gg d$. In Fig.\ref{FigApparentd} we show the scenario where each orbital position is defined by the value $d^{*}$ as a function of the actual distance $d$ and the phase angle $\alpha$.

The relation is obtained by employing the `sine law'. By simple trigonometrical arguments, it can be shown that:
\begin{equation}
\frac{d^{*}}{\sin\varphi}=\frac{D}{\cos\varphi}~~,~~\frac{D}{\sin\alpha}=\frac{d}{\sin\varphi},
\end{equation}
being $\varphi$ the angular separation between EOs for the observer. From the last two relations we can easily find that:
\begin{equation}
d^{*}=\frac{d\sin\alpha}{\cos\left[\arcsin\left(\frac{d}{D}\sin\alpha\right)\right]}\approx d\sin\alpha,
\label{dstar}
\end{equation}
where the last approximation holds in the limit $d\ll D$.

\begin{figure}[!ht]
\includegraphics[width=\linewidth]{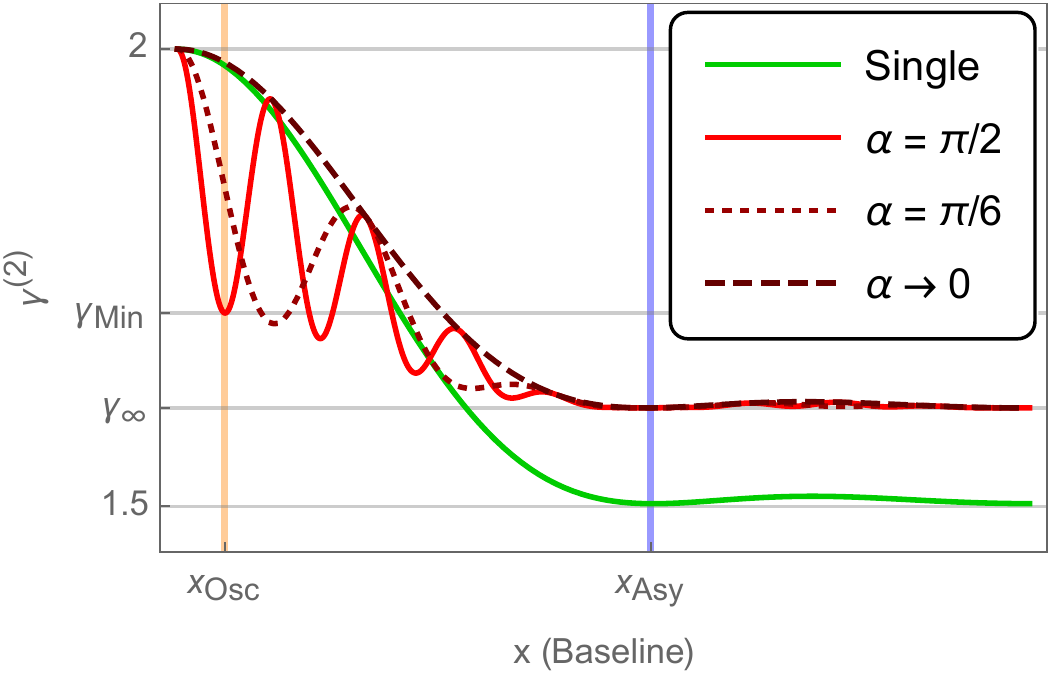}
\caption{\label{FigExampleAlpha} Sketch highlighting how the main features of the second-order coherence change as a function of the baseline (separation between the detectors) $x=x_{1}-x_{2}$ for different values of the phase angle $\alpha$. This corresponds to the replacement $d\rightarrow d^{*}$ given by Eq.(\ref{dstar}) into Eq.(\ref{GammaBinary2nd}). The reddish curves correspond to phase angles $\alpha=0,\pi/6,\pi/2$, respectively, while the green one correspond to the single case. The values $x_{\rm Asy}$ and $x_{\rm Osc}$ correspond to Eqs.(\ref{XBessel}) and (\ref{XOsc}).}
\end{figure}

\begin{figure}[!ht]
\includegraphics[width=\linewidth]{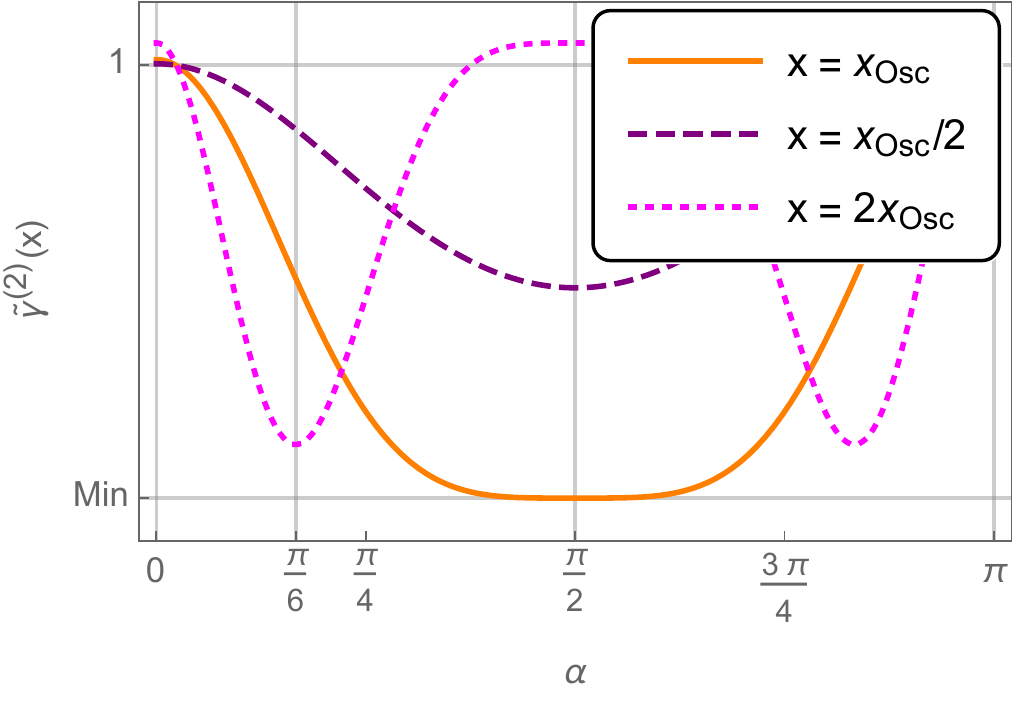}
\caption{\label{FigExampleMinAlpha} Sketch highlighting the behavior of the ratio $\tilde{\gamma}^{(2)}(x)=\gamma^{(2)}(x)/\gamma^{(2)}_{\rm Single}(x)$ as a function of the phase angle $\alpha$ when the two detectors are separated by a distance $x$ equal to $x_{\rm Osc},x_{\rm Osc}/2,2x_{\rm Osc}$. 'Min' corresponds to the value $\gamma_{\rm Min}^{(2)}/\gamma^{(2)}_{\rm Single}(x_{\rm Osc})$.}
\end{figure}

We replace $d\rightarrow d^{*}\approx d\sin\alpha$ in all the expressions of the previous section to include  orbital motion. In Fig.\ref{FigExampleAlpha} we show the second-order coherence of a binary system for different values of the phase angle, $\alpha$.
For $\alpha=0$ (dark red dashed curve), no oscillations appears, even though the curve is different from the single EO found in Fig.\ref{FigVision} because $\gamma_{\infty}^{(2)}\neq 3/2$ in general. For $\alpha\neq 0$, the oscillations take place according to the interference pattern defined by the instantaneous value of the apparent distance $d^{*}$. As the phase angle approaches $\alpha=\pi/2$, more oscillations are contained within the envelope curve. The maximum number of oscillations occurs for the case $\alpha=\pi/2$ (red solid curve), which corresponds to the maximum apparent distance between the two components. For this case, the first minimum of oscillation corresponds to $x_{\rm Osc}$, which is the shortest distance for a minimum to occur. Simultaneously, $\gamma^{(2)}_{\rm Binary}(x=x_{\rm Osc},\alpha=\pi/2)/\gamma^{(2)}_{\rm Single}(x=x_{\rm Osc})$ maximizes the difference between the binary and single cases for every every $x$ and $\alpha$. This is shown in Fig.\ref{FigExampleMinAlpha}, where a normalized second-order coherence $\tilde{\gamma}^{(2)}(x)=\gamma^{(2)}(x)/\gamma^{(2)}_{\rm Single}(x)$ is plotted as a function of $\alpha$.
A measurement at the baseline, $x=x_{\rm Osc}$, gives the best possibility of observing oscillations and, moreover, provides a way to estimate $d$. It also gives the best variation $F_{\rm Osc}$ for determining the rest of the parameters, such as the temperatures or the radii of the objects.

To realize Fig.\ref{FigExampleMinAlpha} experimentally, the light collection time %$\tau_{\rm Col}$ 
required for a single measurement needs to be much smaller than the orbital period $\tau$. Then, as a fair approximation the measurement can be associated to a single position in the orbit (labeled by $\alpha$). For binary stars, the orbital periods can go from days (as the Spica system) to decades (as the Luhman 16 system). The possibility of measurement depends on the particular experimental setup, on how much light is collected and at the same time on the binary system under study.

A crucial aspect about this curve is that it represents a periodic motion. The cases for which the half-period of the orbital motion is accessible experimentally should be separated in two: the systems that can be resolved by complementary methods and, therefore, each measurement can be matched to its corresponding phase angle $\alpha$; and the systems where the latter is not possible since the individual motion of the constituents cannot not be accessed. For the cases where the motion is resolved, the curves of Fig.\ref{FigExampleMinAlpha} should be, in principle, directly obtained. On the other hand, for the cases with not-resolved motion, Fig.\ref{FigExampleMinAlpha} might be still recovered. For this, measurements can be taken as a function of time and by recognizing its periodic structure, the corresponding curve as a function of $\alpha$ could be inferred.

Lastly, if $\tau$ is much larger than the time of experiment and, consequently, measurements for different $\alpha$ are not possible, an estimation of $d$ is still possible by obtaining the curve of Fig.\ref{FigExampleAlpha} provided $\alpha$ is known in advance. Otherwise, those measurements only contribute with an estimation of $d^{*}$, keeping $d$ inaccessible by these means.

In the next sections we apply these results to the study of a paradigmatic case: binary stars.

%%%%%%%%%%%%%%%%%%%%%%%%%%%%%
\section{\label{sec:Luhman16}Application to measurements on binary stars}
%%%%%%%%%%%%%%%%%%%%%%%%%%%%%

We now focus on applying our results to binary stars. These are systems that are vastly found in the Universe. Their classification comes according to the method of observation. The properties of the system condition the method to employ. In this sense, for some binary stars its components can be directly observed. For these cases the optical conditions for observation (associated to the relative motion and relative brightness of the components) are optimal. The two components can be distinguished by direct visualization with appropriate telescopes. Other systems, presenting less advantages for individual visualization of the components require to be explored by other indirect methods. Spectroscopy or the employment of photometry are methods where the presence of a second component is inferred from measurements of Doppler effect or brightness variations by eclipsing orbits, respectively.

Here we aim to exploit the statistics of photon measurements. In principle, the restrictions of the method relate to the observation of the features $\{\gamma_{\rm Min}^{(2)},\gamma_{\infty}^{(2)}\}$ of the second-order coherence summarized in Fig.\ref{FigVision}. As we discussed in Sec.\ref{sec:Measurement}, the measurement strategy depends, on one hand, on the variations $\{F_{\rm Osc},F_{\rm Asy}\}$ for the specific system under observation and, on the other hand, on the baseline possibilities of our intensity interferometer to collect photons of frequencies for which the variations are observable. Lastly, we have to take into account the orbital motion. All in all, depending on the scenario, a complete description of the system could be possible, including the constituents' distance $d$, but also the temperatures and radii of each component $\{R_{\rm A},T_{\rm A},R_{\rm B},T_{\rm B}\}$.

In this work we analyze two cases, the binary stars Luhman 16 and Spica $\alpha$ Vir. The former is a binary brown-dwarf system that shows to be resolved by the South Gemini Observatory in Chile in the visible spectrum. The latter is a spectroscopy binary that is a well-studied case (see Ref.\cite{HanburyBook}). Both constituents stars in Spica are several times larger and hotter than the Sun. Their closeness ($d\sim 0.12{\rm AU}$) and distance ($D\sim 250{\rm ly}$) precludes individual detection. As stated in Ref.\cite{HanburyBook}, intensity interferometry was employed for measuring the radii of the components $\{R_{\rm A},R_{\rm B}\}$ and the distance between them $d$. Here, we also show that with the same technique, the temperatures can be also obtained.

For showing the feasibility of the approach presented here as an alternative measurement method we take the measured values for the parameters of the mentioned binary stars, given in Table \ref{table:LuhmanSpica}.
\begin{table*}
\centering
\begin{tabular}{lccccccccc}
\hline
            & $D$(ly) &  $R_{\rm A}$ & $R_{\rm B}$  & $R_{\rm B}/R_{\rm A}$ & $T_{\rm A}$(K) & $T_{\rm B}$(K) & $d$(AU) & $\tau$ & $m$ \\
\hline
Luhman 16 & 6.51 &  $\sim1.04R_{\rm J}$  &  $\sim0.84R_{\rm J}$ & 0.82 & 1210 & 1350 & 3 & 27.54 years & 10.733 \\
Spica ($\alpha$-Vir) & 250 & $7.47R_{\odot}$ & $3.74R_{\odot}$ & 0.5 & 25300 & 20900 & 0.12 & 4 days & 0.97\\
\hline
\end{tabular}
\caption{\label{table:LuhmanSpica}System parameters of the binary stars Luhman 16 and Spica ($\alpha$-Vir). The distance to the binary star $D$ is given in lightyears (ly). Jupiter's radius is taken as $R_{\rm J}=7.1492~10^{7}{\rm m}$ while the solar radius is $R_{\odot}=6.957~10^{8}{\rm m}$. The distance between the constituents of the binary star $d$ is given in astronomical units (AU). The parameter $\tau$ corresponds to the orbital period, while $m$ stands for the apparent magnitude of the binary stars in the V-band.}
\end{table*}

\begin{figure}[!ht]
\includegraphics[width=\linewidth]{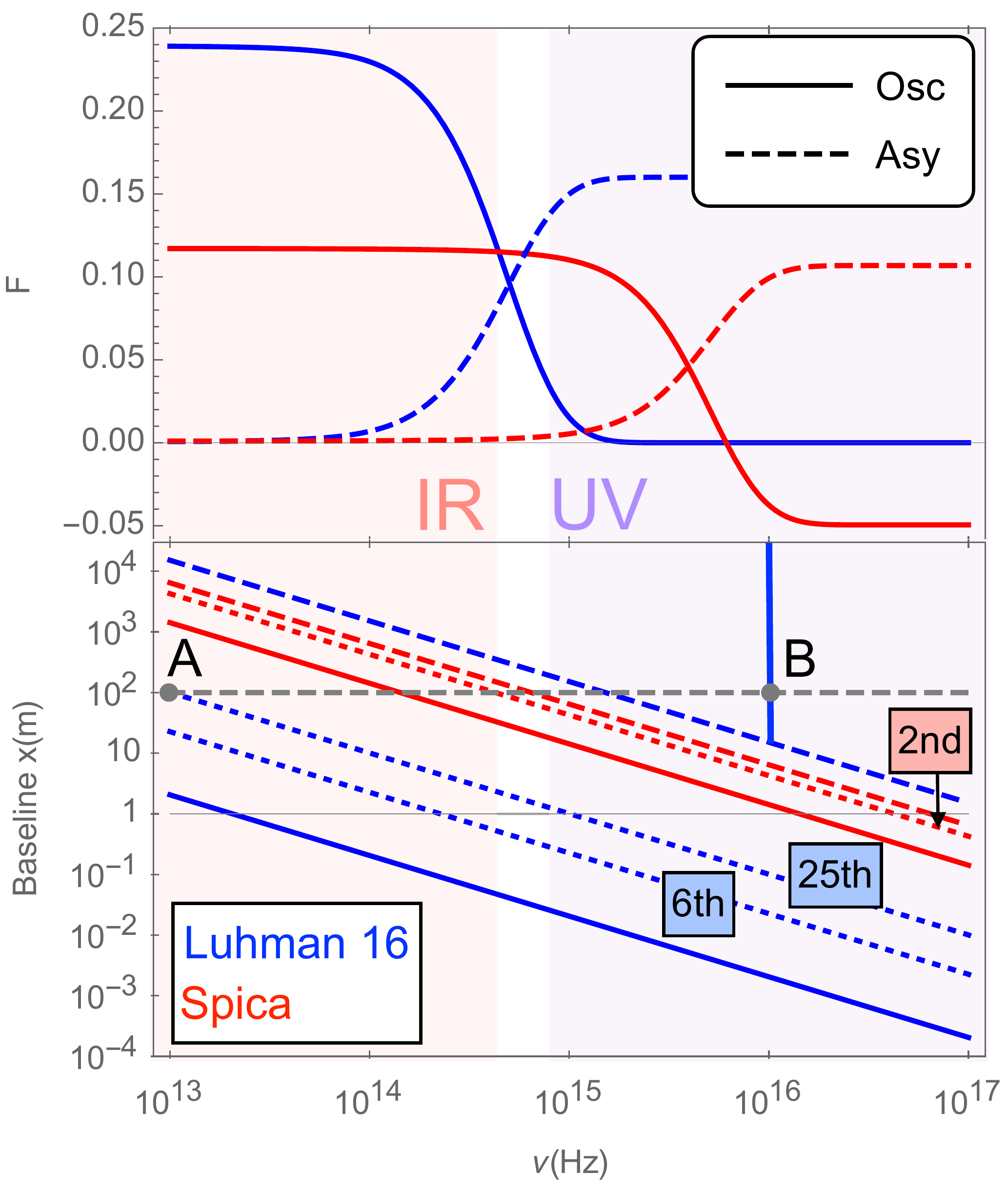}
\caption{\label{FigComparisonLuhmanSpica} \textcolor{blue}{Blue} (\textcolor{red}{red}) curves correspond to Luhman 16 (Spica). Upper: Variations of the oscillation amplitude $F_{\rm Osc}$ (solid curve) and asymptotic value $F_{\rm Asy}$ (dotted curve) as a function of frequency. Lower: Baselines values as a function of the frequency for the first minimum of oscillation [solid curves, $x_{\rm Osc}$ according to Eq.(\ref{XOsc})] and the decay [dotted curves, $x_{\rm Asy}$ according to Eq.(\ref{XBessel})]. The dotted blue (red) lines correspond to the baselines for the 6th and 25th (2nd) minimum, respectively. Point A (B) corresponds to a baseline $x\approx100$m and a frequency $\nu=10$THz ($\nu=10$PHz). The red (violet) shaded region corresponds to the infrared (ultraviolet) frequencies.}
\end{figure}

\begin{figure*}[!ht]
\includegraphics[width=\linewidth]{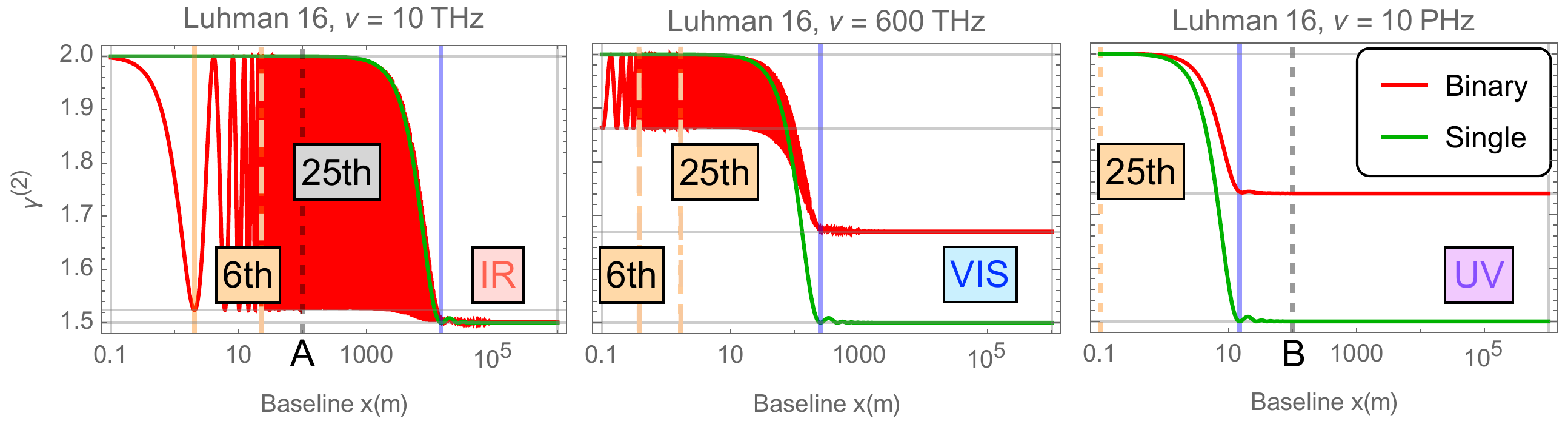}
\caption{\label{FigLuhmanGamma} Second-order coherence $\gamma^{(2)}$ as a function of the baseline $x$ for Luhman 16 for $\alpha=\pi/2$ (at which $d^{*}=d$) and the single object case (taken as the largest constituent alone). The left panel corresponds to a frequency $\nu=10$THz (infrared), the center panel to $\nu=600$THz (visible) and the right one to $\nu=10$PHz (ultraviolet). On every panels, the solid blue vertical lines correspond to the values of the decay baseline $x_{\rm Asy}$ at each frequency [according to Eq.(\ref{XBessel})]. On the left panel, the solid orange vertical line corresponds to the first minimum of oscillation $x_{\rm Osc}$. The dashed orange line corresponds to the position of the minimum $m=6$ as shown in Fig.\ref{FigComparisonLuhmanSpica}. Similarly, the dashed gray curve corresponds to minimum $m=25$, with a baseline $x\approx100$m, corresponding to Point A in Fig.\ref{FigComparisonLuhmanSpica}. Similar notation is shown in the orange dashed lines in the center and right panels. The gray dashed line on the right panel corresponds to one of the possible asymptotic baseline $x\approx100$m, corresponding to Point B in Fig.\ref{FigComparisonLuhmanSpica}. With a baseline $x\approx100$m it is possible to measure both the oscillation amplitude and the asymptotic value by changing the frequency of photocollection.}
\end{figure*}

\begin{figure*}[!ht]
\includegraphics[width=0.9\linewidth]{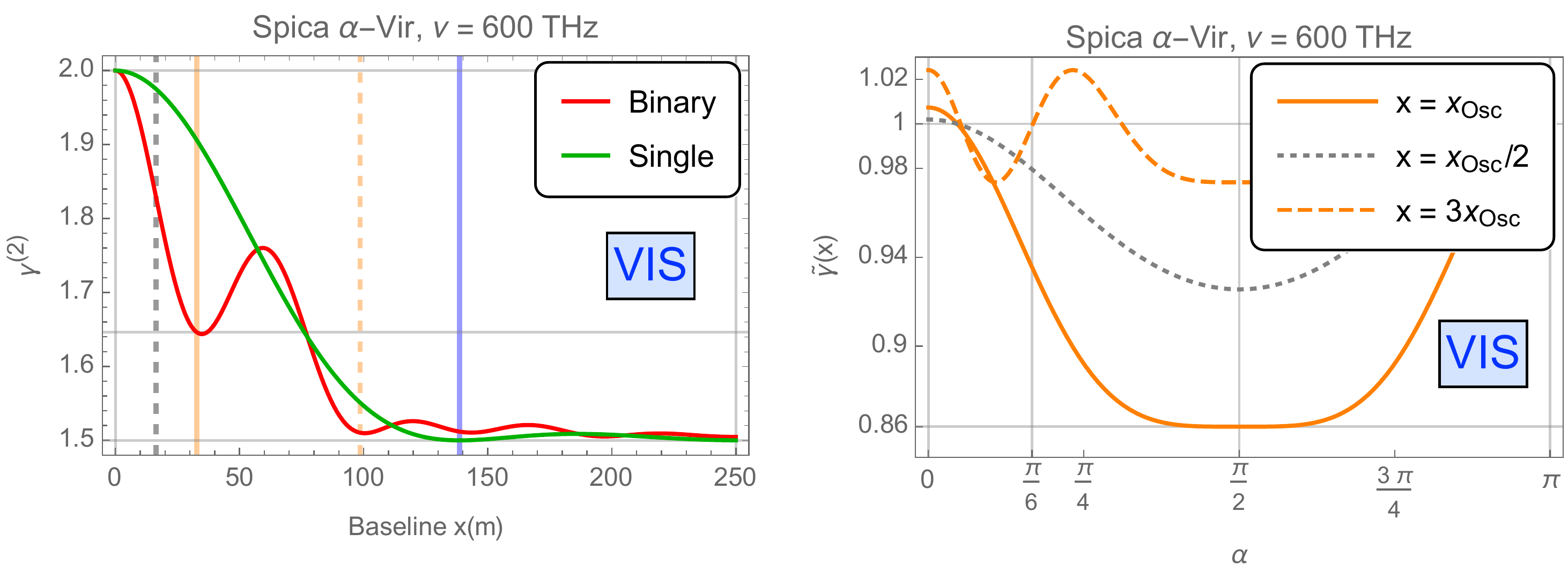}
\caption{\label{FigSpicaGamma} Left panel: Second-order coherence $\gamma^{(2)}$ as a function of the baseline $x$ for Spica $\alpha-$Vir for $\alpha=\pi/2$ (at which $d^{*}=d$) and the single object case (taken as the largest constituent alone) for $\nu=600$THz (in the visible spectrum). The solid orange (blue) line corresponds to the first minimum $x_{\rm Osc}$ (asymptotic value $x_{\rm Asy}$), while the dashed orange corresponds to the 2nd minimum (with a baseline $3x_{\rm Osc}$), which is the only one available. The dashed gray line corresponds to a baseline $x_{\rm Osc}/2$. Right panel: The ratio $\tilde{\gamma}^{(2)}(x)=\gamma^{(2)}(x)/\gamma^{(2)}_{\rm Single}(x)$ as a function of the phase angle $\alpha$ for a frequency $\nu=600$THz in the visible spectrum. Three curves corresponding to the baselines $x=x_{\rm Osc},x_{\rm Osc}/2,3x_{\rm Osc}$, associated to the vertical lines shown on the left panel. The maximum variation is achieved for $x=x_{\rm Osc}$ at the half-angle $\pi/2$, corresponding to the maximum apparent distance between the constituents ($d^{*}=d$). These curves could be obtained for Spica $\alpha-$Vir because of its relatively short period ($\tau=4$ days).}
\end{figure*}

For both systems the size requirement for appreciable variations given in Eq.(\ref{SizeRequirement}) is satisfied. In Fig.\ref{FigComparisonLuhmanSpica} we show the $F_{\rm Osc}$ and $F_{\rm Asy}$, together with the oscillation and decay baselines $\{x_{\rm Osc},x_{\rm Asy}\}$ as functions of the frequency for both binary stars.
As we mentioned on Sec.\ref{sec:Measurement} both plots have to be considered simultaneously in order to determine the best strategy for performing an experimental measurement. In the infrared, for Luhman 16, $F_{\rm Osc}$ reaches 25\% for frequencies approximately up to $100$THz. For the first minimum, the associated baseline is within $\sim 0.3-3{\rm m}$. Nevertheless, given that $x_{\rm Asy}>1$Km, we have that minimums of larger baselines are useful in the same way. As we show, for the minimum corresponding to $m=6$, we have $x_{\rm Osc}\sim 10$m, and for $m=25$ the baselines are closer to 100m. Particularly, for $\nu=10$THz, the oscillation baseline is $x_{\rm Osc}\approx 100$m, corresponding to Point A. In this range of frequencies, $F_{\rm Asy}=0$. These results are also verified in the left panel of Fig.\ref{FigLuhmanGamma}.
For the Spica system, in the infrared we have that $F_{\rm Osc}$ is around 12\%, while $F_{\rm Asy}=0$. In fact, this extends up to the ultraviolet spectrum. The baselines required are between some meters (in the near ultraviolet) to around a kilometer (in the infrared). This agrees with the measurements as commented on Ref.\cite{HanburyBook}, related to oscillation amplitude variations. Unfortunately, in this case, oscillations measurements cannot be complemented by measuring $F_{\rm Asy}$ since there is no variation in this range. Therefore, depending on which property of the system is desired, some prior information might be required. All these aspects can be found also on the left panel of Fig.\ref{FigSpicaGamma}, which is for $\nu=600$THz.
Furthermore, given that the orbital period of Spica is $\tau=4$ days, the right panel shows the analog of Fig.\ref{FigExampleMinAlpha}. We observe that ${\rm Min}=0.86$ for this case, which is only attainable for the first minimum baseline $x_{\rm Osc}$. For Luhman 16 in the visible range the situation is different. Both variations $F_{\rm Osc}$ and $F_{\rm Asy}$ are around 10\%. For oscillations, the baselines are from millimeters to some meters up to $m=25$. For the asymptotic value, the baseline is in the order of hundreds of meters. Thus, for Luhman 16 it is possible to combine both types of measurements by setting two pairs of detectors (in the visible spectrum), with short and large baselines, as can be observed in the center panel of Fig.\ref{FigLuhmanGamma}. In the ultraviolet spectrum, the oscillations disappear while the asymptotic value maximizes $F_{\rm Asy}$. The baselines are below 100m. In particular, for $\nu=1$PHz, a baseline of 100m provides a way to measure the maximum asymptotic value, corresponding to Point B in Fig.\ref{FigComparisonLuhmanSpica} and and the right panel of Fig.\ref{FigLuhmanGamma}. Furthermore, a baseline of 100m provides a way to simultaneously measure the oscillation amplitude and the asymptotic value provided the pair of detectors can measure photons of $\nu=10$THz and $\nu=10$PHz, according to Points A and B. Let us remark that although the latter frequencies are not feasible with actual detectors, the strategy of fixed baseline and variable frequency might be fruitful for some other system. In this sense, we are illustrating all the possible strategies for setting a experiment.

All in all, on general grounds the difference in the baseline scales for each binary stars is mostly given by the difference on the distance to the systems ($D$). As typically happens in optics, a system located further requires a larger baseline. The difference on the oscillation amplitude is due to the similarity of the constituents of each binary star. While the relative sizes ($R_{\rm B}/R_{\rm A}$) are approximately similar, a better situation for Luhman 16 is because the temperatures of the constituents are closer than in the Spica case. The Luhman 16's constituents are more similar to each other than the Spica's, which gives more appreciable oscillations. Furthermore, for Luhman 16 it should be also possible to get an optimal strategy in the visible spectrum, where the system is not resolved, but where the combination of the two features may lead to a full characterization of the system.

Including the orbital motion requires consideration of the orbital period $\tau$. For Luhman 16, the orbital period,  $\tau$, is too large. Therefore, it is not practical to measure the separation distance using photometry observations (in visible). Furthermore, the dependence with the phase angle $\alpha$ (as in Fig.\ref{FigExampleMinAlpha}) may not be possible to observe. As an alternative, the distance between the constituents ($d$) can be estimated from a second-order coherence measurement (as in Fig.\ref{FigExampleAlpha}) if and only if the phase angle is known. Otherwise, only the apparent distance $d^{*}$ can be estimated. In the case of Spica the period is 4 days, so the dependence with the phase angle might be possible to observe, as we mentioned before.

All in all, depending which parameter is to be estimated, the different restrictions for a successful experiment.

%%%%%%%%%%%%%%%%%%%%%%%%%%%%%
\section{\label{sec:Lab}Non-Equilibrium HBT intensity interferometry experiment and feasibility }
%%%%%%%%%%%%%%%%%%%%%%%%%%%%%
In this section, we discuss the practical implementation to measure the second-order coherence function using a typical Hanbury-Brown and Twiss set-up that involves two different optical telescopes (with the desired baseline) equipped with single-photon detectors and timing-cards to measure coincident photon detection events.
The two main conditions that need to be considered are: 

1 -The measurements at both the telescopes should be performed within a time window that allows for the recognition of correlated photon pairs.

2 - Considering the low photon number occupation for the thermal states it is important to estimate the measurement time necessary to get enough coincidence detection events, in other words the signal-to-noise ratio is sufficient enough.

By Measuring second-order correlations as a function of the distance between the telescopes, one measures the correlations between the photons from different points on the wavefront from the stars. Varying the baseline essentially varies the time separation between the points. Thus, based on the coherence time given by $\hbar/T$, where T is the temperature of the object, the correlation function decays. This points to the fact that it is important to perform measurements within a time window that allows for the recognition of correlated photon pairs.  To elucidate, here let us focus a single source such as a mercury vapor lamp or the star Sirius A, as in the original Hanbury-Brown and Twiss experiments. In the experiment in order to detect the correlations it is necessary to measure the two photons coming from the source arrive at the detectors within the coherence time of the source $\tau_{\rm Coh}$. For a mercury vapor lamp the typical coherence time is given by $\hbar/T_{\rm Lamp}$ (being $T_{\rm Lamp}$ the temperature of the lamp) while for a star can be approximately taken as $\tau_{\rm Coh}\sim 10^{-14}{\rm s}$. Following Ref.\cite{Baym}, if the signal is registered over a range within 5-45MHz, the corresponding binning time is around $\tau_{\rm Bin}\sim 10^{-8}{\rm s}$. The probability of observing a pair photons that got correlated during the travel is $\tau_{\rm Bin}/\tau_{\rm Coh}\sim 10^{-6}$. Thus it is necessary to use detectors that are fast enough to enable detection of the correlated pairs. We note that the reasoning and estimations provided hold regardless on the number of stars since the order of magnitude of the coherence time does not change with the properties of the stars.

The second condition aims to address the feasibility of the proposed HBT experiment. This is crucial owing to the fact that the apparent brightness of stars varies over a large range and thus the photon flux reaching the detectors, for example see Table \ref{table:LuhmanSpica}. Furthermore, this is specifically needed for the typical scenarios we address; where objects (binary stars) of different sizes and at different temperatures are involved. The signal-to-noise ratio ($S/N$) required for a given frequency filter band is given as \cite{refId0}:
\begin{equation}
    (S/N)_{\rm RMS}=A\eta F\left[\gamma^{(2)}_{\rm Binary}(x)-3/2%\gamma_{\infty}^{(2)}
    \right]\Delta f^{1/2}\left(\frac{\tau_{\rm exp}}{2}\right)^{1/2},
\end{equation}
where $A$ is the geometric mean of the areas of the two telescopes; $F$ is the photon flux of the sources defined as  number of photons per unit bandwidth, per unit area, and per unit time; $\eta$ corresponds to the quantum efficiency of the detectors; $x$ is the baseline; $\Delta f$ is the electronic bandwidth of the detector plus signal-handling system, and finally $\tau_{\rm exp}$ is the measurement time.

As a feasible experimental scenario, we consider the observation by a pair of telescopes with a radius $R_{\rm Tel}=0.6$m and an efficiency of 0.3 for the photo-detectors. We assume a V-band filter, which corresponds to a wavelength $\lambda=550$nm with a bandwidth $\Delta\lambda=88$nm. The flux for each system is obtained from the apparent magnitudes according to the rule $m=-2.5{\rm log}_{10}(F/F')$ taking $F'$ as the standard reference flux for the V-band. Finally, we consider the electronic bandwidth of the detector to be $1/T_{\rm dead}$, being $T_{\rm dead}$ the dead time of the detector.

For the systems under study, Spica $\alpha$ Vir and Luhman 16 binary stars, the signal-to-noise ratio estimates are encouraging enough to put this experiment proposal into immediate implementation. In the case of the Spica, we have that the total number of photons arriving to each telescope per unit time ($N=\pi R_{\rm Tel}^{2}\eta\Delta\lambda F\lambda/[hc]$, being $c$ the light velocity) is around $4.437\times10^{6}{\rm s}^{-1}$. Since the number saturates typical photo-detectors, we can assume that the collected light is attenuated to a rate $2\times10^{3}{\rm s}^{-1}$. Within these circumstances, to achieve a $(S/N)_{\rm RMS}\approx50$ for the oscillation baseline $x_{\rm Osc}$ an integration time of $\tau_{\rm exp}\approx25{\rm \mu s}$ is required. On the other hand, for Luhman 16, attenuation is not needed. To get $(S/N)_{\rm RMS}\approx35$, the required integration time is $\tau_{\rm exp}\approx10{\rm \mu s}$. We note  a important detail regarding the asymptotic value, taking a baseline $x>x_{\rm Bessel}$, for the Spica we do not notice a appreciable variation in the V-band, see Fig.\ref{FigComparisonLuhmanSpica}. Whilst, for Luhman 16, we get for the signal-to-noise ratio similar numbers as the oscillation amplitude.

%As it was mentioned before, in the present work we are interested in binary systems, formed by two objects (A and B) of different sizes at different temperatures $T_{\rm A,B}$. We consider a scenario where two photo-detectors (1 and 2) are employed for the measurements at the frequency $\omega$. These detectors are characterized by efficiencies $\eta_{\rm 1,2}$ and angular apertures $\Omega_{1,2}$ that condition the collection of photons. The average number of photons $\langle N_{j}\rangle$ arriving to each detector can be written:
%%
%\begin{equation}
%    \langle N_{1,2}\rangle\approx\eta_{1,2}\Omega_{1,2}\left(\bar{n}_{\rm A}+\bar{n}_{\rm B}\right),
%\end{equation}
%%
%where $\bar{n}_{\rm A,B}=1/({\rm Exp}[\hbar\omega/(k_{\rm B}T_{\rm A,B})]-1)$ is the mean photon number at temperature $T_{\rm A,B}$. Nevertheless, for 

%%%%%%%%%%%%%%%%%%%%%%%%%%%%%
\section{\label{sec:Conclusions}Conclusions}
%%%%%%%%%%%%%%%%%%%%%%%%%%%%%

In this work we have entered the study of spatial coherence on photon statistics particularly applied to nonequilibrium configurations of sources. For this, we have developed a general formalism for calculating the electric field correlation functions as functional derivatives of a suitable generating functional. Implementing the P-function representation, we have focused on situations involving two extended spherical objects (formed by a continuous of point-sources) at different temperatures. Specifically, we have calculated the second-order coherence associated to coincident counts of photons at two different points of space. As in the Hanbury-Brown and Twiss experiment of intensity interferometer, both pair of objects and detectors lie on the same plane.

As general results, we demonstrated that the second-order coherence of binary scenarios are mainly characterized by two features: 1- oscillations as a function of the baseline $x$ (distance between the detectors); and 2- a long-baseline asymptotic value. On one hand, the latter is characterized by a value $\gamma_{\infty}^{(2)}$ [Eq.(\ref{GammaLD})] and is found for baselines satisfying $x>x_{\rm Asy}$ [Eq.(\ref{XBessel})]. On the other hand, the former is a reminiscence of the effect of photon bunching. It can be exploited by the employment of the minimums of oscillation. The first one, characterized by a value $\gamma_{\rm Min}^{(2)}$ [Eq.(\ref{GammaMin})] and a baseline $x_{\rm Osc}$ [Eq.(\ref{XOsc})], is in principle the best of them for extracting information of the system. However, for scenarios where $x_{\rm Osc}\ll x_{\rm Asy}$, the minimums with larger baselines ($(2m-1)x_{\rm Osc}$) can give the same information as the first since $\gamma^{(2)}_{\rm Binary}([2m-1]x_{\rm Osc})\approx\gamma_{\rm Min}^{(2)}$.

We then analyzed how to define the best measurement strategy according to different possible scenarios. As both features depend in different manners on the frequency of the photons collected, the surface ratio of the objects and both temperatures, we show what are the crucial aspects defining the interplay between the variations of each feature $\{F_{\rm Osc},F_{\rm Asy}\}$ and the baselines. We show that competition and complementarity of the features can be found in different scenarios. While the oscillations are found to be enhanced when the two sources are similar (similar sizes and photon mean numbers), the asymptotic value is enhanced for similar sizes but greater thermal imbalances (different photon mean numbers). Also, as the photon mean numbers depends on frequency of the photons collected, we analyzed how to combine measurements in different regions of the spectrum. The fact that temperatures are part of the possible estimations provided by the approach opens a new branch of application for intensity interferometry and Hanbury-Brown and Twiss experiments.

Motivated by scenarios where one of objects is orbiting another one, we included orbital motion in our calculations through a simplified model of circular orbits. We explored the impact of this motion on the second-order coherence. Depending on the prior information about the system and the orbital period itself, the measurements of $\gamma_{\rm Min}^{(2)}$ could lead to an estimation of the actual ($d$) or, alternatively, just to the apparent distance between the objects ($d^{*}$).

We applied our model and results to the case of two binary stars, Luhman 16 and Spica $\alpha$ Vir. Their choice was based on the fact that their different properties demonstrate the broad application of our approach to actual scenarios.

From the experimental side, we have analyzed some aspects on the feasibility, estimating the integration time required for measuring the features in the V-band. An estimation of the signal-to-noise ratio for the measurements of photo-coincidences showed promising numbers for the binary stars considered. For measurements of the oscillation amplitude we obtained an integration time on the order of a few tenths of microseconds for signal-to-noise ratios of 35-50. For the asymptotic value, in the V-band, it is only possible for the Luhman 16 case, obtaining similar values. For the Spica case the variation in the asymptotic value is not appreciable.

All in all, we believe that the present work is contributing to take current available methodologies one step further, including nonequilibrium configurations. This gives the chance to a better characterization of the systems under study. In this sense, let us remark that although we have applied our results to astrophysical systems, the same formalism may be useful for scenarios involving microscopical sources out of thermal equilibrium.

\begin{acknowledgments}
This work is funded by Defense Advanced Research Projects Agency (DARPA). The opinions and findings are those of the authors and should not be interpreted as representing the official views or policies of the Department of Defense or the U.S. Government.
\end{acknowledgments}

%%%%%%%%%%%%%%%%%%%%%%

\appendix

%%%%%%%%%%%%%%%%%%%%%%
\section{A generating functional for the correlation functions}\label{App:GeneratingFunctional}
%%%%%%%%%%%%%%%%%%%%%%

This appendix is devoted to show how the correlation functions for the electric field operator can be obtained from a generating functional.

We start by considering the concept of spatial coherence, which relates to measurements of photons in different points of space. These measurements are associated to the quantum correlations of the electric field operator at different points of space. In our case, we start by defining the $k$th-order correlation function by following Ref.\cite{Milonni}:
\begin{eqnarray}
&&G^{(k)}_{\alpha_{1}...\alpha_{k},\alpha'_{1}...\alpha'_{k}}(x_{1},...,x_{k};x'_{1},...,x'_{k})=\nonumber\\
&&=\left\langle : \hat{E}^{(-)}_{\alpha_{1}}(x_{1})\hat{E}^{(+)}_{\alpha'_{1}}(x'_{1})...\hat{E}^{(-)}_{\alpha_{k}}(x_{k})\hat{E}^{(+)}_{\alpha'_{k}}(x'_{k}):\right\rangle\\
&&=\left\langle \hat{E}^{(-)}_{\alpha_{1}}(x_{1})...\hat{E}^{(-)}_{\alpha_{k}}(x_{k})\hat{E}^{(+)}_{\alpha'_{1}}(x'_{1})...\hat{E}^{(+)}_{\alpha'_{k}}(x'_{k})\right\rangle,\nonumber
\end{eqnarray}
%
%or shortly,
%
%\begin{eqnarray}
%G^{(k)}_{\alpha_{1}...\alpha_{k},\alpha'_{1}...\alpha'_{k}}(x_{1},...,x_{k};x'_{1},...,x'_{k})&=&\left\langle : \hat{\mathbf{E}}^{(-)}(x_{1})\hat{\mathbf{E}}^{(+)}(x'_{1})...\hat{\mathbf{E}}^{(-)}(x_{k})\hat{\mathbf{E}}^{(+)}(x'_{k}):\right\rangle\\
%&=&\left\langle \hat{\mathbf{E}}^{(-)}(x_{1})...\hat{\mathbf{E}}^{(-)}(x_{k})\hat{\mathbf{E}}^{(+)}(x'_{1})...\hat{\mathbf{E}}^{(+)}(x'_{k})\right\rangle,\nonumber
%\end{eqnarray}
%
where $:~:$ stands for the normal product, the coordinates are $x_{i}\equiv (\mathbf{x}_{i},t_{i})$ while $\{\alpha_{i}\}$ stands for the components of the electric field operators.

In our case, we will be interested in the case where these functions are connected to measurements of photons in different points of space. In this sense, we are interested in the case where $x_{i}=x'_{i}$ and $\alpha_{i}=\alpha'_{i}$. Then, the notation simplifies to:
\begin{equation}
G^{(k)}_{\alpha_{1}...\alpha_{k}}(x_{1},...,x_{k})\equiv G^{(k)}_{[\alpha_{1}...\alpha_{k},\alpha_{1}...\alpha_{k}]}(x_{1},...,x_{k};x_{1},...,x_{k}),
\label{GkSpa-Pol-Coherence}
\end{equation}
where the subscripts in squared brackets $[\alpha_{1}...\alpha_{k},\alpha_{1}...\alpha_{k}]$ imply that there is no sum although the subscripts are repeated.

Notice that these functions will be measuring correlations at different spacetime points between different polarizations of the electric field. In that sense, if we want to go further and only explore correlations at different spatial points, then we should set $\alpha_{i}=\alpha$ and $t_{i}=t$ for every $i$. Thus, we get correlation functions that can be written as:
\begin{eqnarray}
G^{(k)}(\mathbf{x}_{1},...,\mathbf{x}_{k},t)&\equiv&\left\langle \hat{E}^{(-)}(\mathbf{x}_{1},t)...\hat{E}^{(-)}(\mathbf{x}_{k},t)\right.\label{CorrelationKSpatial}\\
&&\left.\times\hat{E}^{(+)}(\mathbf{x}_{1},t)...\hat{E}^{(+)}(\mathbf{x}_{k},t)\right\rangle\nonumber\\
&=&G^{(k)}_{[\alpha...\alpha,\alpha...\alpha]}(\mathbf{x}_{1},t,...,\mathbf{x}_{k},t;\mathbf{x}_{1},t,...,\mathbf{x}_{k},t),\nonumber
\end{eqnarray}
where we have ommited the polarization subscripts for simplicity.

Observing the definitions of the quantum correlations functions for spatial coherence on Eq.(\ref{GkSpa-Pol-Coherence}), we turn to analyze the functional defined as:
\begin{widetext}
\begin{equation}
\mathcal{Z}\left[a_{jk}(\mathbf{x},\tau)\right]\equiv\left\langle\mathcal{T}:{\rm Exp}\left(\int d\mathbf{x}\int_{0}^{t_{\rm f}}d\tau~a_{jk}(\mathbf{x},\tau)\hat{E}^{(-)}_{j}(\mathbf{x},\tau)\hat{E}^{(+)}_{k}(\mathbf{x},\tau)\right):\right\rangle,
\label{GeneratingFunctional}
\end{equation}
where $a_{jk}(\mathbf{x},\tau)$ stands for, in principle, an arbitrary distribution defined in all the spacetime points where the electric field operators are spanned. The time $t_{\rm f}$ in principle is arbitrary.

For the particular case of analyzing the spatial coherence of the electric field, we set $a_{jk}(\mathbf{x},\tau)\equiv a_{jk}(\mathbf{x})\delta(\tau-t)$ with $0<t<t_{\rm f}$. Thus, the functional $\mathcal{Z}$ reduces to:
\begin{equation}
Z_{\rm SC}\left[a_{jk}(\mathbf{x})\right]=\left\langle\mathcal{T}:{\rm Exp}\left(\int d\mathbf{x}~a_{jk}(\mathbf{x})\hat{E}^{(-)}_{j}(\mathbf{x},t)\hat{E}^{(+)}_{k}(\mathbf{x},t)\right):\right\rangle.
\label{GeneratingFunctionalSC}
\end{equation}

If, in addition, we consider the case for which a single component of the electric field is analyzed (i.e. $\hat{\mathbf{E}}=\hat{E}\mathbf{e}$), the generating functional $Z_{\rm SC}$ simplifies to:
\begin{equation}
Z\left[a(\mathbf{x})\right]=\left\langle\mathcal{T}:{\rm Exp}\left(\int d\mathbf{x}~a(\mathbf{x})\hat{E}^{(-)}(\mathbf{x},t)\hat{E}^{(+)}(\mathbf{x},t)\right):\right\rangle.
\label{GeneratingFunctionalParallel}
\end{equation}

Then, we immediately notice that the functional derivative of the functional $Z$ with respect to $a$ at $\mathbf{x}_{1}$ reads:
\begin{equation}
\frac{\delta Z}{\delta a(\mathbf{x}_{1})}=\left\langle\mathcal{T}:\left[\int d\mathbf{x}~\frac{\delta a(\mathbf{x})}{\delta a(\mathbf{x}_{1})}\hat{E}^{(-)}(\mathbf{x},t)\hat{E}^{(+)}(\mathbf{x},t)\right]{\rm Exp}\left(\int d\mathbf{x}~a(\mathbf{x})\hat{E}^{(-)}(\mathbf{x},t)\hat{E}^{(+)}(\mathbf{x},t)\right):\right\rangle.
\end{equation}

By noticing that :
\begin{equation}
\frac{\delta a(\mathbf{x})}{\delta a(\mathbf{x}_{1})}=\delta(\mathbf{x}-\mathbf{x}_{1}),
\end{equation}
we immediately get that:
\begin{equation}
\frac{\delta Z}{\delta a(\mathbf{x}_{1})}=\left\langle\mathcal{T}:\left[\hat{E}^{(-)}(\mathbf{x}_{1},t)\hat{E}^{(+)}(\mathbf{x}_{1},t)\right]{\rm Exp}\left(\int d\mathbf{x}~a(\mathbf{x})\hat{E}^{(-)}(\mathbf{x},t)\hat{E}^{(+)}(\mathbf{x},t)\right):\right\rangle.
\end{equation}
\end{widetext}

%At this point, we notice the fundamental property that the first-order correlation function ($k=1$) corresponding to Eq.(\ref{GkSpa-Pol-Coherence}) can be written as:
%
%\begin{equation}
%G^{(1)}_{\alpha}(\mathbf{x}_{1},t)=\left.\frac{\delta Z_{\rm SC}}{\delta a_{\alpha\alpha}(\mathbf{x}_{1})}\right|_{a=0}.
%\end{equation}

In the same way, we can easily extend this to every order, noticing finally that:
\begin{equation}
G^{(k)}(\mathbf{x}_{1},...,\mathbf{x}_{k},t)=\left.\frac{\delta^{k} Z}{\delta a(\mathbf{x}_{1})...\delta a(\mathbf{x}_{k})}\right|_{a=0},
\end{equation}
which corresponds to Eq.(\ref{QuantumCorrelationKDerivatives}).

This shows that we can obtain the correlation functions of arbitrary order of the electric field for studying the spatial coherence [Eq.(\ref{CorrelationKSpatial})] from the generating functional of Eq.(\ref{GeneratingFunctionalParallel}). As developed, this procedure is general and the information about the particular source configuration and its states is encoded in the electric field operator and the expectation value.

%%%%%%%%%%%%%%%%%%%%%%
\section{Calculation of the generating functional for two thermal point-sources}\label{App:GeneratingFunctionalTwoThermalSources}
%%%%%%%%%%%%%%%%%%%%%%

This appendix shows how the generating functional of Eq.(\ref{ZGeneratingEE}) is calculated for the scenario of two thermal point-sources to finally obtain the result of Eq.(\ref{ZTwoSourcesFunctional}).

We start by implementing the two-sources state for single modes in astrophysical scenarios, summarized in Eqs.(\ref{StateGeneral}) and (\ref{ElectricSingleMode}), for calculating the expectation value of the r.h.s. of Eq.(\ref{ZGeneratingEE}), provided the thermal P-functions characterizing the state of each source [Eq.(\ref{P-FunctionThermal})]. Then, we have that the generating functional reads:
\begin{eqnarray}
&&Z\left[a(\mathbf{x})\right]
=\\
&&=\frac{1}{\pi^{2}\bar{n}_{1}\bar{n}_{2}}\int\int e^{-\frac{|v_{1}|^{2}}{\bar{n}_{1}}-\frac{|v_{2}|^{2}}{\bar{n}_{2}}+\sum_{l,m=1}^{2}A_{ml}v_{l}^{*}v_{m}}d^{2}v_{1}d^{2}v_{2},\nonumber
\end{eqnarray}
with the coefficients given by:
\begin{equation}
A_{ml}\equiv\mathcal{E}_{0}^{2}%\frac{}{\epsilon_{0}}
\int d\mathbf{x}~e^{i\frac{\omega}{c}\Delta R_{ml}(\mathbf{x})}a(\mathbf{x}),
\end{equation}
where $\Delta R_{ij}(\mathbf{x})=R_{i}(\mathbf{x})-R_{j}(\mathbf{x})$ corresponds to the difference in the optical path traveled by the radiation coming from each source, so $R_{i}(\mathbf{x})=\mathbf{n}_{i}\cdot\mathbf{r}_{i}+ct_{i}$.

Moreover, the functional derivative of these coefficients reads:
\begin{equation}
\frac{\delta A_{ml}}{\delta a(\mathbf{x}')}\equiv\mathcal{E}_{0}^{2}%\frac{}{\epsilon_{0}}
e^{i\frac{\omega}{c}\Delta R_{ml}(\mathbf{x}')}.
\end{equation}

The resulting integral consists in a multi-dimensional Gaussian integral as long as $1/\bar{n}_{l}-A_{ll}>0$ for $l=1,2$. This is the case for the two possible evaluations of $a$ that we are considering, so the integrals converge. Splitting each variable $v_{1,2}$ in real and imaginary parts (with $d^{2}v=(d{\rm Re}[v])d({\rm Im}[v])$), the integral reads:
\begin{eqnarray}
&&\int\int e^{-\frac{|v_{1}|^{2}}{\bar{n}_{1}}-\frac{|v_{2}|^{2}}{\bar{n}_{2}}+\sum_{l,m=1}^{2}A_{ml}v_{l}^{*}v_{m}}d^{2}v_{1}d^{2}v_{2}=\\
&&=\int_{\mathbb{R}^{4}}e^{-\frac{1}{2}\mathbf{v}^{T}\cdot\mathbb{A}\cdot\mathbf{v}}d\mathbf{v}=\sqrt{\frac{(2\pi)^{4}}{\det\mathbb{A}}},\nonumber
\end{eqnarray}
with $\mathbf{v}=({\rm Re}[v_{1}],{\rm Im}[v_{1}],{\rm Re}[v_{2}],{\rm Im}[v_{2}])$ and $d\mathbf{v}=d^{2}v_{1}d^{2}v_{2}$, while the matrix is given by:
\begin{widetext}
\begin{equation}
\mathbb{A}=
\begin{pmatrix}
2\left(\frac{1}{\bar{n}_{1}}-A_{11}\right) & 0 & -2{\rm Re}(A_{12}) & -2{\rm Im}(A_{12})\\
0 & 2\left(\frac{1}{\bar{n}_{1}}-A_{11}\right) & 2{\rm Im}(A_{12}) & -2{\rm Re}(A_{12})\\
-2{\rm Re}(A_{12}) & 2{\rm Im}(A_{12}) & 2\left(\frac{1}{\bar{n}_{2}}-A_{22}\right) & 0\\
-2{\rm Im}(A_{12}) & -2{\rm Re}(A_{12}) & 0 & 2\left(\frac{1}{\bar{n}_{2}}-A_{22}\right)
\end{pmatrix},
\end{equation}
\end{widetext}
which determinant results:
\begin{equation}
\det\mathbb{A}=\left[\frac{4\left[\left(1-A_{11}\bar{n}_{1}\right)\left(1-A_{22}\bar{n}_{2}\right)-\bar{n}_{1}\bar{n}_{2}\left|A_{12}\right|^{2}\right]}{\bar{n}_{1}\bar{n}_{2}}\right]^{2}.
\end{equation}

Thus, we have that the generating functional reads:
\begin{equation}
Z\left[a(\mathbf{x})\right]
=\frac{1}{\left[\left(1-A_{11}\bar{n}_{1}\right)\left(1-A_{22}\bar{n}_{2}\right)-\bar{n}_{1}\bar{n}_{2}\left|A_{12}\right|^{2}\right]},
\end{equation}
which corresponds to Eq.(\ref{ZTwoSourcesFunctional}).

%%%%%%%%%%%%%%%%%%%%%%
\section{Calculation of the first two correlation functions}\label{App:CorrelationsTwoThermalSources}
%%%%%%%%%%%%%%%%%%%%%%

This appendix is devoted to show the calculation of the first two correlation functions given in Eqs.(\ref{G1TwoThermalSources}) and (\ref{G2TwoThermalSources}) from the generating functional of Eq.(\ref{ZTwoSourcesFunctional}). The scenario of interest consists in two thermal point-sources at different temperature. The involved distances are typical of an astrophysical scenario, having that the distance between the sources ($d\equiv|\mathbf{r}_{1}-\mathbf{r}_{2}|$) and the distance between all the observation points $|\mathbf{x}_{i}-\mathbf{x}_{j}|$ for all the pairs of observation points ($i,j=1,...,k$) are smaller than the distance $D$ between the sources and the observation points, satisfying $D\gg d\gg|\mathbf{x}_{i}-\mathbf{x}_{j}|$.

At a first point it is important to write down the functional derivatives of the generating functional given in Eq.(\ref{ZTwoSourcesFunctional}):
\begin{widetext}
\begin{equation}
\frac{\delta Z}{\delta a(\mathbf{x}_{1})}=\left[\frac{\delta A_{11}}{\delta a(\mathbf{x}_{1})}\bar{n}_{1}\left(1-A_{22}\bar{n}_{2}\right)+\left(1-A_{11}\bar{n}_{1}\right)\frac{\delta A_{22}}{\delta a(\mathbf{x}_{1})}\bar{n}_{2}+\bar{n}_{1}\bar{n}_{2}\left(\frac{\delta A_{12}^{*}}{\delta a(\mathbf{x}_{1})}A_{12}+A_{12}^{*}\frac{\delta A_{12}}{\delta a(\mathbf{x}_{1})}\right)\right]\left(Z\left[a(\mathbf{x})\right]\right)^{2}.
\end{equation}
\begin{eqnarray}
\frac{\delta^{2}Z}{\delta a(\mathbf{x}_{1})\delta a(\mathbf{x}_{2})}&=&\bar{n}_{1}\bar{n}_{2}\left[-\frac{\delta A_{11}}{\delta a(\mathbf{x}_{1})}\frac{\delta A_{22}}{\delta a(\mathbf{x}_{2})}-\frac{\delta A_{11}}{\delta a(\mathbf{x}_{2})}\frac{\delta A_{22}}{\delta a(\mathbf{x}_{1})}+\frac{\delta A_{12}^{*}}{\delta a(\mathbf{x}_{1})}\frac{\delta A_{12}}{\delta a(\mathbf{x}_{2})}+\frac{\delta A_{12}^{*}}{\delta a(\mathbf{x}_{2})}\frac{\delta A_{12}}{\delta a(\mathbf{x}_{1})}\right]\left(Z\left[a(\mathbf{x})\right]\right)^{2}\nonumber\\
&+&2\left[\frac{\delta A_{11}}{\delta a(\mathbf{x}_{1})}\bar{n}_{1}\left(1-A_{22}\bar{n}_{2}\right)+\left(1-A_{11}\bar{n}_{1}\right)\frac{\delta A_{22}}{\delta a(\mathbf{x}_{1})}\bar{n}_{2}+\bar{n}_{1}\bar{n}_{2}\left(\frac{\delta A_{12}^{*}}{\delta a(\mathbf{x}_{1})}A_{12}+A_{12}^{*}\frac{\delta A_{12}}{\delta a(\mathbf{x}_{1})}\right)\right]\\
&\times&\left[\frac{\delta A_{11}}{\delta a(\mathbf{x}_{2})}\bar{n}_{1}\left(1-A_{22}\bar{n}_{2}\right)+\left(1-A_{11}\bar{n}_{1}\right)\frac{\delta A_{22}}{\delta a(\mathbf{x}_{2})}\bar{n}_{2}+\bar{n}_{1}\bar{n}_{2}\left(\frac{\delta A_{12}^{*}}{\delta a(\mathbf{x}_{2})}A_{12}+A_{12}^{*}\frac{\delta A_{12}}{\delta a(\mathbf{x}_{2})}\right)\right]\left(Z\left[a(\mathbf{x})\right]\right)^{3}.\nonumber
\end{eqnarray}

Apart from this, we can calculate the first two quantum correlation functions from the generating functional, obtaining:
\begin{equation}
G^{(1)}=\bar{n}_{1}\left.\frac{\delta A_{11}}{\delta a(\mathbf{x}_{1})}\right|_{a=0}+\bar{n}_{2}\left.\frac{\delta A_{22}}{\delta a(\mathbf{x}_{1})}\right|_{a=0}=\mathcal{E}_{0}^{2}%\frac{}{\epsilon_{0}}
\left(\bar{n}_{1}+\bar{n}_{2}\right),
\label{G1TwoThermalSourcesApp}
\end{equation}
\begin{eqnarray}
G^{(2)}(\mathbf{x}_{1},\mathbf{x}_{2})&=&2\bar{n}_{1}^{2}\left.\frac{\delta A_{11}}{\delta a(\mathbf{x}_{1})}\right|_{a=0}\left.\frac{\delta A_{11}}{\delta a(\mathbf{x}_{2})}\right|_{a=0}+2\bar{n}_{2}^{2}\left.\frac{\delta A_{22}}{\delta a(\mathbf{x}_{1})}\right|_{a=0}\left.\frac{\delta A_{22}}{\delta a(\mathbf{x}_{2})}\right|_{a=0}+\bar{n}_{1}\bar{n}_{2}\left[\left.\frac{\delta A_{22}}{\delta a(\mathbf{x}_{2})}\right|_{a=0}\left.\frac{\delta A_{11}}{\delta a(\mathbf{x}_{1})}\right|_{a=0}\right.\\
&+&\left.\left.\frac{\delta A_{11}}{\delta a(\mathbf{x}_{2})}\right|_{a=0}\left.\frac{\delta A_{22}}{\delta a(\mathbf{x}_{1})}\right|_{a=0}+\left.\frac{\delta A_{12}}{\delta a(\mathbf{x}_{1})}\right|_{a=0}\left.\frac{\delta A_{12}^{*}}{\delta a(\mathbf{x}_{2})}\right|_{a=0}+\left.\frac{\delta A_{12}^{*}}{\delta a(\mathbf{x}_{1})}\right|_{a=0}\left.\frac{\delta A_{12}}{\delta a(\mathbf{x}_{2})}\right|_{a=0}\right]\nonumber\\
&=&2\mathcal{E}_{0}^{4}%\frac{}{\epsilon_{0}^{2}}
\left(\bar{n}_{1}^{2}+\bar{n}_{2}^{2}+\bar{n}_{1}\bar{n}_{2}\left[1+\cos\left(\frac{\omega}{c}\left[\Delta R_{12}(\mathbf{x}_{1})-\Delta R_{12}(\mathbf{x}_{2})\right]\right)\right]\right).\nonumber
\end{eqnarray}
%
%$2*n1^{2}*A11^{(0,1)}[0,0]*A11^{(1,0)}[0,0]+2*n2^{2}*A22^{(1,0)}[0,0]*A22^{(0,1)}[0,0]+n1*n2(A22^{(0,1)}[0,0]*A11^{(1,0)}[0,0]+A11^{(0,1)}[0,0]*A22^{(1,0)}[0,0]+ConA12^{(0,1)}[0,0]*A12^{(1,0)}[0,0]+A12^{(0,1)}[0,0]*ConA12^{(1,0)}[0,0])$
\end{widetext}

Notice that in the case of thermal equilibrium ($\bar{n}_{1}=\bar{n}_{2}=\bar{n}$), the expression directly agrees with the one obtained in Eq.(4.4.18) of Ref.\cite{Scully}. Moreover, Eq.(\ref{G1TwoThermalSourcesApp}) corresponds directly to Eq.(\ref{G1TwoThermalSources}).

Considering the integrations involved in the present work and the typical astrophysical scenarios mentioned, the positions are given by:
\begin{equation}
\mathbf{r}_{1,2}=d_{1,2}\check{x}+D\check{y}+z_{1,2}\check{z},
\label{PositionsStar}
\end{equation}

With the mentioned approximations, we immediately obtain:
\begin{eqnarray}
\left|\mathbf{x}_{i}-\mathbf{r}_{j}\right|&=&\sqrt{(x_{i}-d_{j})^{2}+z_{j}^{2}+D^{2}}\label{DistantSourcesApprox}\\
&=&D\sqrt{1+\left(\frac{x_{i}-d_{j}}{D}\right)^{2}+\left(\frac{z_{j}}{D}\right)^{2}}\nonumber\\
&\approx& D\left[1+\frac{1}{2}\left(\frac{x_{i}-d_{j}}{D}\right)^{2}+\frac{1}{2}\left(\frac{z_{j}}{D}\right)^{2}+...\right].\nonumber
\end{eqnarray}

Within this regime, the second-order coherence reads:
\begin{eqnarray}
G^{(2)}(x_{1}-x_{2})&\approx&2\mathcal{E}_{0}^{4}%\frac{}{\epsilon_{0}^{2}}
\left(\bar{n}_{1}^{2}+\bar{n}_{2}^{2}+\bar{n}_{1}\bar{n}_{2}\right.\\
&\times&\left.\left[1+\cos\left(\frac{\omega [d_{1}-d_{2}]}{cD}[x_{1}-x_{2}]\right)\right]\right).\nonumber
\end{eqnarray}

Notice that the coordinates $z_{1,2}$ make no effect in the second-order coherence at the end since the observation points are on the $x-$axis.

Notice that the last expression corresponds exactly to Eq.(\ref{G2TwoThermalSources}).

Finally, we proved the expressions for the first two correlation functions for two thermal sources of different temperatures in an astrophysical scenario.

%%%%%%%%%%%%%%%%%%%%%%
\section{Calculation of the second-order coherence for an astrophysical binary system}\label{App:SecondOrderCoherenceBinary}
%%%%%%%%%%%%%%%%%%%%%%

This appendix shows the calculation of the second-order coherence for a binary system (a pair of AEOs), starting from Eq.(\ref{Binary2ndCoherence}) and arriving to Eqs.(\ref{GammaBinary2nd}), (\ref{GammaLD}) and (\ref{GammaIJ}). The scenario consists in two spherical objects of radii $R_{\rm A,B}$ and at different temperatures $T_{\rm A,B}$ (associated to mean photon numbers $\bar{n}_{\rm A,B}$, respectively). As we mentioned in the main text, a crucial approximation is that the surfaces $S_{\rm A,B}$ are taken as discs of radii $R_{\rm A,B}$. Since $D\gg |\mathbf{r}_{1}-\mathbf{r}_{2}|\gg|\mathbf{x}_{i}-\mathbf{x}_{j}|$, the curvature of the objects with respect to the observers is negligible so the integration can be taken over the flat cross section. Thus, the surfaces $S_{\rm A,B}\approx\pi R_{\rm A,B}^{2}$. Finally, we consider the separation between the centers of each constituent as $d$.

In general, the second-order coherence for the binary system results from summing the second-order coherences of all the pair of points, given by Eq.(\ref{Binary2ndCoherence}):
\begin{eqnarray}
\gamma_{\rm Binary}^{(2)}(\mathbf{x}_{1},\mathbf{x}_{2})&=&\frac{1}{(S_{\rm A}+S_{\rm B})^{2}}\int_{S_{\rm A}\cup S_{\rm B}}dS_{1}\int_{S_{\rm A}\cup S_{\rm B}}dS_{2}\nonumber\\
&\times&\gamma^{(2)}(\mathbf{x}_{1},\mathbf{x}_{2},\mathbf{r}_{1},\mathbf{r}_{2}).
\end{eqnarray}

Considering that the second-order coherence for a pair of points is given by Eq.(\ref{Gamma22S}), we have that the second-order coherence for the binary system reads:
\begin{widetext}
\begin{equation}
\gamma_{\rm Binary}^{(2)}(x_{1}-x_{2})=\frac{2}{(S_{\rm A}+S_{\rm B})^{2}}\sum_{i,j={\rm A,B}}\int_{S_{i}}dS_{1}\int_{S_{j}}dS_{2}\left(1+\frac{\bar{n}_{i}\bar{n}_{j}}{\left(\bar{n}_{i}+\bar{n}_{j}\right)^{2}}\left[\cos\left(\frac{\omega(d_{1}-d_{2})}{cD}(x_{1}-x_{2})\right)-1\right]\right)%\\
%&=&\frac{2}{(S_{\rm A}+S_{\rm B})^{2}}\sum_{i,j={\rm A,B}}\left[\left(1-\frac{\bar{n}_{i}\bar{n}_{j}}{\left(\bar{n}_{i}+\bar{n}_{j}\right)^{2}}\right)S_{i}S_{j}+\frac{\bar{n}_{i}\bar{n}_{j}}{\left(\bar{n}_{i}+\bar{n}_{j}\right)^{2}}\int_{S_{i}}dS_{1}\int_{S_{j}}dS_{2}\cos\left(\frac{\omega(d_{1}-d_{2})}{cD}(x_{1}-x_{2})\right)\right].\nonumber
\end{equation}

For the surface integrations we should consider that the binary system presents component A centered in the origin of the $x-z$ plane, while component B is centered around a point on the $x-$axis located at a distance $d$. Then, for the integrations over the points of the component A, we have $\int_{S_{\rm A}}dS_{j}=\int_{0}^{R_{\rm A}}dr_{j}\int_{0}^{2\pi}d\theta_{j}~r_{j}$, so $d_{j}=r_{j}\cos\theta_{j}$. On the other hand, for the integrations over the points of the component B, we have $\int_{S_{\rm B}}dS_{j}=\int_{0}^{R_{\rm B}}dr_{j}\int_{0}^{2\pi}d\theta_{j}~r_{j}$ and $d_{j}=d+r_{j}\cos\theta_{j}$. In general, the positions can be written as $d_{j}=\delta_{j,{\rm B}}d+r_{j}\cos\theta_{j}$. Using that $\cos(a-b)=\cos(a)\cos(b)+\sin(a)\sin(b)$ and the integrals:
\begin{equation}
\int_{0}^{R}dr\int_{0}^{2\pi}d\theta r\cos\left[\frac{\omega r\cos\theta}{cD}(x_{1}-x_{2})\right]=\frac{2\pi cDR}{\omega(x_{1}-x_{2})}J_{1}\left(\frac{\omega R}{cD}(x_{1}-x_{2})\right),
\end{equation}
\begin{equation}
\int_{0}^{R}dr\int_{0}^{2\pi}d\theta r\sin\left[\frac{\omega r\cos\theta}{cD}(x_{1}-x_{2})\right]=0,
\end{equation}
then, we have:
\begin{eqnarray}
\int_{S_{i}}dS_{1}\int_{S_{j}}dS_{2}\cos\left(\frac{\omega(d_{1}-d_{2})}{cD}(x_{1}-x_{2})\right)&=&\left[\delta_{i,{\rm A}}\delta_{j,{\rm A}}+\delta_{i,{\rm B}}\delta_{j,{\rm B}}+\cos\left(\frac{\omega d}{cD}(x_{1}-x_{2})\right)\left(\delta_{i,{\rm A}}\delta_{j,{\rm B}}+\delta_{i,{\rm B}}\delta_{j,{\rm A}}\right)\right]\\
&\times&\frac{2\pi cDR_{i}}{\omega(x_{1}-x_{2})}J_{1}\left(\frac{\omega R_{i}}{cD}(x_{1}-x_{2})\right)\frac{2\pi cDR_{j}}{\omega(x_{1}-x_{2})}J_{1}\left(\frac{\omega R_{j}}{cD}(x_{1}-x_{2})\right).\nonumber
\end{eqnarray}
\end{widetext}

Therefore, for a binary system we have:
\begin{eqnarray}
\gamma_{\rm Binary}^{(2)}(x_{1}-x_{2})&=&\gamma_{\infty}^{(2)}+\frac{1}{(1+s)^{2}}\left[\gamma_{\rm AA}^{(2)}(x_{1}-x_{2})\right.\nonumber\\
&+&\left.s^{2}\gamma_{\rm BB}^{(2)}(x_{1}-x_{2})
+\frac{8\mathcal{N}s}{(1+\mathcal{N})^{2}}\right.\\
&\times&\left.\cos\left(\frac{\omega d}{cD}[x_{1}-x_{2}]\right)\gamma_{\rm AB}^{(2)}(x_{1}-x_{2})\right],\nonumber
\end{eqnarray}
with each contribution given by:
\begin{equation}
\gamma_{\infty}^{(2)}=\frac{3}{2}+\frac{s}{(1+s)^{2}}\frac{(1-\mathcal{N})^{2}}{(1+\mathcal{N})^{2}},
\end{equation}
\begin{eqnarray}
\gamma_{\rm ij}^{(2)}(x_{1}-x_{2})&=&\frac{2}{R_{i}R_{j}}\left[\frac{cD}{\omega(x_{1}-x_{2})}\right]^{2}\\
&\times&J_{1}\left(\frac{\omega R_{i}}{cD}[x_{1}-x_{2}]\right)J_{1}\left(\frac{\omega R_{j}}{cD}[x_{1}-x_{2}]\right),\nonumber
\end{eqnarray}
where $s=S_{\rm B}/S_{\rm A}=(R_{\rm B}/R_{\rm A})^{2}$ is the surface ratio and $\mathcal{N}=\bar{n}_{\rm B}/\bar{n}_{\rm A}$ the ratio between the mean photon numbers at each temperature. Notice that the last equations correspond to Eqs.(\ref{GammaBinary2nd}), (\ref{GammaLD}) and (\ref{GammaIJ}).

\nocite{*}

\bibliography{apssamp}% Produces the bibliography via BibTeX.

\providecommand{\noopsort}[1]{}\providecommand{\singleletter}[1]{#1}%
\begin{thebibliography}{30}%
\makeatletter
\providecommand \@ifxundefined [1]{%
 \@ifx{#1\undefined}
}%
\providecommand \@ifnum [1]{%
 \ifnum #1\expandafter \@firstoftwo
 \else \expandafter \@secondoftwo
 \fi
}%
\providecommand \@ifx [1]{%
 \ifx #1\expandafter \@firstoftwo
 \else \expandafter \@secondoftwo
 \fi
}%
\providecommand \natexlab [1]{#1}%
\providecommand \enquote  [1]{``#1''}%
\providecommand \bibnamefont  [1]{#1}%
\providecommand \bibfnamefont [1]{#1}%
\providecommand \citenamefont [1]{#1}%
\providecommand \href@noop [0]{\@secondoftwo}%
\providecommand \href [0]{\begingroup \@sanitize@url \@href}%
\providecommand \@href[1]{\@@startlink{#1}\@@href}%
\providecommand \@@href[1]{\endgroup#1\@@endlink}%
\providecommand \@sanitize@url [0]{\catcode `\\12\catcode `\$12\catcode
  `\&12\catcode `\#12\catcode `\^12\catcode `\_12\catcode `\%12\relax}%
\providecommand \@@startlink[1]{}%
\providecommand \@@endlink[0]{}%
\providecommand \url  [0]{\begingroup\@sanitize@url \@url }%
\providecommand \@url [1]{\endgroup\@href {#1}{\urlprefix }}%
\providecommand \urlprefix  [0]{URL }%
\providecommand \Eprint [0]{\href }%
\providecommand \doibase [0]{https://doi.org/}%
\providecommand \selectlanguage [0]{\@gobble}%
\providecommand \bibinfo  [0]{\@secondoftwo}%
\providecommand \bibfield  [0]{\@secondoftwo}%
\providecommand \translation [1]{[#1]}%
\providecommand \BibitemOpen [0]{}%
\providecommand \bibitemStop [0]{}%
\providecommand \bibitemNoStop [0]{.\EOS\space}%
\providecommand \EOS [0]{\spacefactor3000\relax}%
\providecommand \BibitemShut  [1]{\csname bibitem#1\endcsname}%
\let\auto@bib@innerbib\@empty
%</preamble>
\bibitem [{\citenamefont {Brown}\ and\ \citenamefont {Twiss}(2013)}]{HBT1956}%
  \BibitemOpen
  \bibfield  {author} {\bibinfo {author} {\bibfnamefont {R.~H.}\ \bibnamefont
  {Brown}}\ and\ \bibinfo {author} {\bibfnamefont {R.~Q.}\ \bibnamefont
  {Twiss}},\ }\bibfield  {title} {\bibinfo {title} {2. a test of a new type of
  stellar interferometer on sirius},\ }in\ \href@noop {} {\emph {\bibinfo
  {booktitle} {A Source Book in Astronomy and Astrophysics, 1900--1975}}}\
  (\bibinfo  {publisher} {Harvard University Press},\ \bibinfo {year} {2013})\
  pp.\ \bibinfo {pages} {8--12}\BibitemShut {NoStop}%
\bibitem [{\citenamefont {Glauber}(1963)}]{Glauber}%
  \BibitemOpen
  \bibfield  {author} {\bibinfo {author} {\bibfnamefont {R.~J.}\ \bibnamefont
  {Glauber}},\ }\bibfield  {title} {\bibinfo {title} {Coherent and incoherent
  states of the radiation field},\ }\href@noop {} {\bibfield  {journal}
  {\bibinfo  {journal} {Physical Review}\ }\textbf {\bibinfo {volume} {131}},\
  \bibinfo {pages} {2766} (\bibinfo {year} {1963})}\BibitemShut {NoStop}%
\bibitem [{\citenamefont {Sudarshan}(1963)}]{Sudarshan}%
  \BibitemOpen
  \bibfield  {author} {\bibinfo {author} {\bibfnamefont {E.~C.~G.}\
  \bibnamefont {Sudarshan}},\ }\bibfield  {title} {\bibinfo {title}
  {Equivalence of semiclassical and quantum mechanical descriptions of
  statistical light beams},\ }\href
  {https://doi.org/10.1103/PhysRevLett.10.277} {\bibfield  {journal} {\bibinfo
  {journal} {Phys. Rev. Lett.}\ }\textbf {\bibinfo {volume} {10}},\ \bibinfo
  {pages} {277} (\bibinfo {year} {1963})}\BibitemShut {NoStop}%
\bibitem [{\citenamefont {{Grangier}}\ \emph {et~al.}(1986)\citenamefont
  {{Grangier}}, \citenamefont {{Roger}}, \citenamefont {{Aspect}},
  \citenamefont {{Heidmann}},\ and\ \citenamefont {{Reynaud}}}]{Antibunching}%
  \BibitemOpen
  \bibfield  {author} {\bibinfo {author} {\bibfnamefont {P.}~\bibnamefont
  {{Grangier}}}, \bibinfo {author} {\bibfnamefont {G.}~\bibnamefont {{Roger}}},
  \bibinfo {author} {\bibfnamefont {A.}~\bibnamefont {{Aspect}}}, \bibinfo
  {author} {\bibfnamefont {A.}~\bibnamefont {{Heidmann}}},\ and\ \bibinfo
  {author} {\bibfnamefont {S.}~\bibnamefont {{Reynaud}}},\ }\bibfield  {title}
  {\bibinfo {title} {{Observation of photon antibunching in phase-matched
  multiatom resonance fluorescence}},\ }\href
  {https://doi.org/10.1103/PhysRevLett.57.687} {\bibfield  {journal} {\bibinfo
  {journal} {\prl}\ }\textbf {\bibinfo {volume} {57}},\ \bibinfo {pages} {687}
  (\bibinfo {year} {1986})}\BibitemShut {NoStop}%
\bibitem [{\citenamefont {{Sattler}}\ and\ \citenamefont
  {{Hartfuss}}(1993)}]{Plasma}%
  \BibitemOpen
  \bibfield  {author} {\bibinfo {author} {\bibfnamefont {S.}~\bibnamefont
  {{Sattler}}}\ and\ \bibinfo {author} {\bibfnamefont {H.~J.}\ \bibnamefont
  {{Hartfuss}}},\ }\bibfield  {title} {\bibinfo {title} {{Intensity
  interferometry for measurement of electron temperature fluctuations in fusion
  plasmas}},\ }\href {https://doi.org/10.1088/0741-3335/35/9/016} {\bibfield
  {journal} {\bibinfo  {journal} {Plasma Physics and Controlled Fusion}\
  }\textbf {\bibinfo {volume} {35}},\ \bibinfo {pages} {1285} (\bibinfo {year}
  {1993})}\BibitemShut {NoStop}%
\bibitem [{\citenamefont {Baym}(1998)}]{Baym}%
  \BibitemOpen
  \bibfield  {author} {\bibinfo {author} {\bibfnamefont {G.}~\bibnamefont
  {Baym}},\ }\bibfield  {title} {\bibinfo {title} {The physics of hanbury
  brown--twiss intensity interferometry: from stars to nuclear collisions},\
  }\href@noop {} {\bibfield  {journal} {\bibinfo  {journal} {arXiv preprint
  arXiv:nucl-th/9804026.}\ } (\bibinfo {year} {1998})}\BibitemShut {NoStop}%
\bibitem [{\citenamefont {Giovannini}(2011)}]{Cosmology}%
  \BibitemOpen
  \bibfield  {author} {\bibinfo {author} {\bibfnamefont {M.}~\bibnamefont
  {Giovannini}},\ }\bibfield  {title} {\bibinfo {title} {Hanbury brown--twiss
  interferometry and second-order correlations of inflaton quanta},\ }\href
  {https://doi.org/10.1103/PhysRevD.83.023515} {\bibfield  {journal} {\bibinfo
  {journal} {Phys. Rev. D}\ }\textbf {\bibinfo {volume} {83}},\ \bibinfo
  {pages} {023515} (\bibinfo {year} {2011})}\BibitemShut {NoStop}%
\bibitem [{\citenamefont {Csernai}\ \emph {et~al.}(2015)\citenamefont
  {Csernai}, \citenamefont {Hatlen},\ and\ \citenamefont
  {Zschocke}}]{DifferentialHBT}%
  \BibitemOpen
  \bibfield  {author} {\bibinfo {author} {\bibfnamefont {L.~P.}\ \bibnamefont
  {Csernai}}, \bibinfo {author} {\bibfnamefont {E.~S.}\ \bibnamefont
  {Hatlen}},\ and\ \bibinfo {author} {\bibfnamefont {S.}~\bibnamefont
  {Zschocke}},\ }\bibfield  {title} {\bibinfo {title} {Differential hbt method
  for binary stars},\ }\href@noop {} {\bibfield  {journal} {\bibinfo  {journal}
  {arXiv preprint arXiv:1505.07342}\ } (\bibinfo {year} {2015})}\BibitemShut
  {NoStop}%
\bibitem [{\citenamefont {{Cohen}}\ \emph {et~al.}(2015)\citenamefont
  {{Cohen}}, \citenamefont {{Meenehan}}, \citenamefont {{Maccabe}},
  \citenamefont {{Gr{\"o}blacher}}, \citenamefont {{Safavi-Naeini}},
  \citenamefont {{Marsili}}, \citenamefont {{Shaw}},\ and\ \citenamefont
  {{Painter}}}]{Nanoresonator}%
  \BibitemOpen
  \bibfield  {author} {\bibinfo {author} {\bibfnamefont {J.~D.}\ \bibnamefont
  {{Cohen}}}, \bibinfo {author} {\bibfnamefont {S.~M.}\ \bibnamefont
  {{Meenehan}}}, \bibinfo {author} {\bibfnamefont {G.~S.}\ \bibnamefont
  {{Maccabe}}}, \bibinfo {author} {\bibfnamefont {S.}~\bibnamefont
  {{Gr{\"o}blacher}}}, \bibinfo {author} {\bibfnamefont {A.~H.}\ \bibnamefont
  {{Safavi-Naeini}}}, \bibinfo {author} {\bibfnamefont {F.}~\bibnamefont
  {{Marsili}}}, \bibinfo {author} {\bibfnamefont {M.~D.}\ \bibnamefont
  {{Shaw}}},\ and\ \bibinfo {author} {\bibfnamefont {O.}~\bibnamefont
  {{Painter}}},\ }\bibfield  {title} {\bibinfo {title} {{Phonon counting and
  intensity interferometry of a nanomechanical resonator}},\ }\href
  {https://doi.org/10.1038/nature14349} {\bibfield  {journal} {\bibinfo
  {journal} {\nat}\ }\textbf {\bibinfo {volume} {520}},\ \bibinfo {pages} {522}
  (\bibinfo {year} {2015})},\ \Eprint {https://arxiv.org/abs/1410.1047}
  {arXiv:1410.1047 [quant-ph]} \BibitemShut {NoStop}%
\bibitem [{\citenamefont {Kanno}\ and\ \citenamefont
  {Soda}(2019)}]{NonclassicalGW}%
  \BibitemOpen
  \bibfield  {author} {\bibinfo {author} {\bibfnamefont {S.}~\bibnamefont
  {Kanno}}\ and\ \bibinfo {author} {\bibfnamefont {J.}~\bibnamefont {Soda}},\
  }\bibfield  {title} {\bibinfo {title} {Detecting nonclassical primordial
  gravitational waves with hanbury-brown--twiss interferometry},\ }\href
  {https://doi.org/10.1103/PhysRevD.99.084010} {\bibfield  {journal} {\bibinfo
  {journal} {Phys. Rev. D}\ }\textbf {\bibinfo {volume} {99}},\ \bibinfo
  {pages} {084010} (\bibinfo {year} {2019})}\BibitemShut {NoStop}%
\bibitem [{\citenamefont {Rai}\ \emph {et~al.}(2021)\citenamefont {Rai},
  \citenamefont {Basak},\ and\ \citenamefont {Saha}}]{SimulationsBinary}%
  \BibitemOpen
  \bibfield  {author} {\bibinfo {author} {\bibfnamefont {K.~N.}\ \bibnamefont
  {Rai}}, \bibinfo {author} {\bibfnamefont {S.}~\bibnamefont {Basak}},\ and\
  \bibinfo {author} {\bibfnamefont {P.}~\bibnamefont {Saha}},\ }\bibfield
  {title} {\bibinfo {title} {Radius measurement in binary stars: simulations of
  intensity interferometry},\ }\href {https://doi.org/10.1093/mnras/stab2391}
  {\bibfield  {journal} {\bibinfo  {journal} {Monthly Notices of the Royal
  Astronomical Society}\ }\textbf {\bibinfo {volume} {507}},\ \bibinfo {pages}
  {2813–2824} (\bibinfo {year} {2021})}\BibitemShut {NoStop}%
\bibitem [{\citenamefont {Bojer}\ \emph {et~al.}(2021)\citenamefont {Bojer},
  \citenamefont {Huang}, \citenamefont {Karl}, \citenamefont {Richter},
  \citenamefont {Kok},\ and\ \citenamefont {von Zanthier}}]{VonZanthier2021}%
  \BibitemOpen
  \bibfield  {author} {\bibinfo {author} {\bibfnamefont {M.}~\bibnamefont
  {Bojer}}, \bibinfo {author} {\bibfnamefont {Z.}~\bibnamefont {Huang}},
  \bibinfo {author} {\bibfnamefont {S.}~\bibnamefont {Karl}}, \bibinfo {author}
  {\bibfnamefont {S.}~\bibnamefont {Richter}}, \bibinfo {author} {\bibfnamefont
  {P.}~\bibnamefont {Kok}},\ and\ \bibinfo {author} {\bibfnamefont
  {J.}~\bibnamefont {von Zanthier}},\ }\href@noop {} {\bibinfo {title} {A
  quantitative comparison of amplitude versus intensity interferometry for
  astronomy}} (\bibinfo {year} {2021}),\ \Eprint
  {https://arxiv.org/abs/2106.05640} {arXiv:2106.05640 [astro-ph.IM]}
  \BibitemShut {NoStop}%
\bibitem [{\citenamefont {Thiel}\ \emph {et~al.}(2009)\citenamefont {Thiel},
  \citenamefont {Bastin}, \citenamefont {von Zanthier},\ and\ \citenamefont
  {Agarwal}}]{thiel2009sub}%
  \BibitemOpen
  \bibfield  {author} {\bibinfo {author} {\bibfnamefont {C.}~\bibnamefont
  {Thiel}}, \bibinfo {author} {\bibfnamefont {T.}~\bibnamefont {Bastin}},
  \bibinfo {author} {\bibfnamefont {J.}~\bibnamefont {von Zanthier}},\ and\
  \bibinfo {author} {\bibfnamefont {G.~S.}\ \bibnamefont {Agarwal}},\
  }\bibfield  {title} {\bibinfo {title} {Sub-rayleigh quantum imaging using
  single-photon sources},\ }\href@noop {} {\bibfield  {journal} {\bibinfo
  {journal} {Physical Review A}\ }\textbf {\bibinfo {volume} {80}},\ \bibinfo
  {pages} {013820} (\bibinfo {year} {2009})}\BibitemShut {NoStop}%
\bibitem [{\citenamefont {Cui}\ \emph {et~al.}(2013)\citenamefont {Cui},
  \citenamefont {Sun}, \citenamefont {Chen}, \citenamefont {Gong},\ and\
  \citenamefont {Guo}}]{cui2013quantum}%
  \BibitemOpen
  \bibfield  {author} {\bibinfo {author} {\bibfnamefont {J.-M.}\ \bibnamefont
  {Cui}}, \bibinfo {author} {\bibfnamefont {F.-W.}\ \bibnamefont {Sun}},
  \bibinfo {author} {\bibfnamefont {X.-D.}\ \bibnamefont {Chen}}, \bibinfo
  {author} {\bibfnamefont {Z.-J.}\ \bibnamefont {Gong}},\ and\ \bibinfo
  {author} {\bibfnamefont {G.-C.}\ \bibnamefont {Guo}},\ }\bibfield  {title}
  {\bibinfo {title} {Quantum statistical imaging of particles without
  restriction of the diffraction limit},\ }\href@noop {} {\bibfield  {journal}
  {\bibinfo  {journal} {Physical review letters}\ }\textbf {\bibinfo {volume}
  {110}},\ \bibinfo {pages} {153901} (\bibinfo {year} {2013})}\BibitemShut
  {NoStop}%
\bibitem [{\citenamefont {Israel}\ \emph {et~al.}(2017)\citenamefont {Israel},
  \citenamefont {Tenne}, \citenamefont {Oron},\ and\ \citenamefont
  {Silberberg}}]{israel2017quantum}%
  \BibitemOpen
  \bibfield  {author} {\bibinfo {author} {\bibfnamefont {Y.}~\bibnamefont
  {Israel}}, \bibinfo {author} {\bibfnamefont {R.}~\bibnamefont {Tenne}},
  \bibinfo {author} {\bibfnamefont {D.}~\bibnamefont {Oron}},\ and\ \bibinfo
  {author} {\bibfnamefont {Y.}~\bibnamefont {Silberberg}},\ }\bibfield  {title}
  {\bibinfo {title} {Quantum correlation enhanced super-resolution localization
  microscopy enabled by a fibre bundle camera},\ }\href@noop {} {\bibfield
  {journal} {\bibinfo  {journal} {Nature communications}\ }\textbf {\bibinfo
  {volume} {8}},\ \bibinfo {pages} {1} (\bibinfo {year} {2017})}\BibitemShut
  {NoStop}%
\bibitem [{\citenamefont {Classen}\ \emph {et~al.}(2017)\citenamefont
  {Classen}, \citenamefont {von Zanthier}, \citenamefont {Scully},\ and\
  \citenamefont {Agarwal}}]{classen2017superresolution}%
  \BibitemOpen
  \bibfield  {author} {\bibinfo {author} {\bibfnamefont {A.}~\bibnamefont
  {Classen}}, \bibinfo {author} {\bibfnamefont {J.}~\bibnamefont {von
  Zanthier}}, \bibinfo {author} {\bibfnamefont {M.~O.}\ \bibnamefont
  {Scully}},\ and\ \bibinfo {author} {\bibfnamefont {G.~S.}\ \bibnamefont
  {Agarwal}},\ }\bibfield  {title} {\bibinfo {title} {Superresolution via
  structured illumination quantum correlation microscopy},\ }\href@noop {}
  {\bibfield  {journal} {\bibinfo  {journal} {Optica}\ }\textbf {\bibinfo
  {volume} {4}},\ \bibinfo {pages} {580} (\bibinfo {year} {2017})}\BibitemShut
  {NoStop}%
\bibitem [{\citenamefont {{Tenne}}\ \emph {et~al.}(2019)\citenamefont
  {{Tenne}}, \citenamefont {{Rossman}}, \citenamefont {{Rephael}},
  \citenamefont {{Israel}}, \citenamefont {{Krupinski-Ptaszek}}, \citenamefont
  {{Lapkiewicz}}, \citenamefont {{Silberberg}},\ and\ \citenamefont
  {{Oron}}}]{tenne2019super}%
  \BibitemOpen
  \bibfield  {author} {\bibinfo {author} {\bibfnamefont {R.}~\bibnamefont
  {{Tenne}}}, \bibinfo {author} {\bibfnamefont {U.}~\bibnamefont {{Rossman}}},
  \bibinfo {author} {\bibfnamefont {B.}~\bibnamefont {{Rephael}}}, \bibinfo
  {author} {\bibfnamefont {Y.}~\bibnamefont {{Israel}}}, \bibinfo {author}
  {\bibfnamefont {A.}~\bibnamefont {{Krupinski-Ptaszek}}}, \bibinfo {author}
  {\bibfnamefont {R.}~\bibnamefont {{Lapkiewicz}}}, \bibinfo {author}
  {\bibfnamefont {Y.}~\bibnamefont {{Silberberg}}},\ and\ \bibinfo {author}
  {\bibfnamefont {D.}~\bibnamefont {{Oron}}},\ }\bibfield  {title} {\bibinfo
  {title} {{Super-resolution enhancement by quantum image scanning
  microscopy}},\ }\href {https://doi.org/10.1038/s41566-018-0324-z} {\bibfield
  {journal} {\bibinfo  {journal} {Nature Photonics}\ }\textbf {\bibinfo
  {volume} {13}},\ \bibinfo {pages} {116} (\bibinfo {year} {2019})},\ \Eprint
  {https://arxiv.org/abs/1806.07661} {arXiv:1806.07661 [physics.optics]}
  \BibitemShut {NoStop}%
\bibitem [{\citenamefont {Forbes}\ and\ \citenamefont
  {Rodriguez-Fajardo}(2019)}]{forbes2019super}%
  \BibitemOpen
  \bibfield  {author} {\bibinfo {author} {\bibfnamefont {A.}~\bibnamefont
  {Forbes}}\ and\ \bibinfo {author} {\bibfnamefont {V.}~\bibnamefont
  {Rodriguez-Fajardo}},\ }\bibfield  {title} {\bibinfo {title}
  {Super-resolution with quantum light},\ }\href@noop {} {\bibfield  {journal}
  {\bibinfo  {journal} {Nature Photonics}\ }\textbf {\bibinfo {volume} {13}},\
  \bibinfo {pages} {76} (\bibinfo {year} {2019})}\BibitemShut {NoStop}%
\bibitem [{\citenamefont {Gottesman}\ \emph {et~al.}(2012)\citenamefont
  {Gottesman}, \citenamefont {Jennewein},\ and\ \citenamefont {Croke}}]{Croke}%
  \BibitemOpen
  \bibfield  {author} {\bibinfo {author} {\bibfnamefont {D.}~\bibnamefont
  {Gottesman}}, \bibinfo {author} {\bibfnamefont {T.}~\bibnamefont
  {Jennewein}},\ and\ \bibinfo {author} {\bibfnamefont {S.}~\bibnamefont
  {Croke}},\ }\bibfield  {title} {\bibinfo {title} {Longer-baseline telescopes
  using quantum repeaters},\ }\href
  {https://doi.org/10.1103/PhysRevLett.109.070503} {\bibfield  {journal}
  {\bibinfo  {journal} {Phys. Rev. Lett.}\ }\textbf {\bibinfo {volume} {109}},\
  \bibinfo {pages} {070503} (\bibinfo {year} {2012})}\BibitemShut {NoStop}%
\bibitem [{\citenamefont {Tsang}\ \emph {et~al.}(2016)\citenamefont {Tsang},
  \citenamefont {Nair},\ and\ \citenamefont {Lu}}]{tsang2016quantum}%
  \BibitemOpen
  \bibfield  {author} {\bibinfo {author} {\bibfnamefont {M.}~\bibnamefont
  {Tsang}}, \bibinfo {author} {\bibfnamefont {R.}~\bibnamefont {Nair}},\ and\
  \bibinfo {author} {\bibfnamefont {X.-M.}\ \bibnamefont {Lu}},\ }\bibfield
  {title} {\bibinfo {title} {Quantum theory of superresolution for two
  incoherent optical point sources},\ }\href@noop {} {\bibfield  {journal}
  {\bibinfo  {journal} {Physical Review X}\ }\textbf {\bibinfo {volume} {6}},\
  \bibinfo {pages} {031033} (\bibinfo {year} {2016})}\BibitemShut {NoStop}%
\bibitem [{\citenamefont {Bao}\ \emph {et~al.}(2021)\citenamefont {Bao},
  \citenamefont {Choi}, \citenamefont {Aggarwal},\ and\ \citenamefont
  {Jacob}}]{BaoQAI}%
  \BibitemOpen
  \bibfield  {author} {\bibinfo {author} {\bibfnamefont {F.}~\bibnamefont
  {Bao}}, \bibinfo {author} {\bibfnamefont {H.}~\bibnamefont {Choi}}, \bibinfo
  {author} {\bibfnamefont {V.}~\bibnamefont {Aggarwal}},\ and\ \bibinfo
  {author} {\bibfnamefont {Z.}~\bibnamefont {Jacob}},\ }\bibfield  {title}
  {\bibinfo {title} {Quantum-accelerated imaging of n stars},\ }\href
  {https://doi.org/10.1364/OL.430404} {\bibfield  {journal} {\bibinfo
  {journal} {Opt. Lett.}\ }\textbf {\bibinfo {volume} {46}},\ \bibinfo {pages}
  {3045} (\bibinfo {year} {2021})}\BibitemShut {NoStop}%
\bibitem [{\citenamefont {Gerry}\ \emph {et~al.}(2005)\citenamefont {Gerry},
  \citenamefont {Knight},\ and\ \citenamefont
  {Knight}}]{gerry2005introductory}%
  \BibitemOpen
  \bibfield  {author} {\bibinfo {author} {\bibfnamefont {C.}~\bibnamefont
  {Gerry}}, \bibinfo {author} {\bibfnamefont {P.}~\bibnamefont {Knight}},\ and\
  \bibinfo {author} {\bibfnamefont {P.}~\bibnamefont {Knight}},\ }\href
  {https://books.google.com/books?id=CgByyoBJJwgC} {\emph {\bibinfo {title}
  {Introductory Quantum Optics}}}\ (\bibinfo  {publisher} {Cambridge University
  Press},\ \bibinfo {year} {2005})\BibitemShut {NoStop}%
\bibitem [{\citenamefont {Scully}\ and\ \citenamefont
  {Zubairy}(1999)}]{Scully}%
  \BibitemOpen
  \bibfield  {author} {\bibinfo {author} {\bibfnamefont {M.~O.}\ \bibnamefont
  {Scully}}\ and\ \bibinfo {author} {\bibfnamefont {M.~S.}\ \bibnamefont
  {Zubairy}},\ }\href@noop {} {\bibinfo {title} {Quantum optics}} (\bibinfo
  {year} {1999})\BibitemShut {NoStop}%
\bibitem [{\citenamefont {Milonni}(2013)}]{Milonni}%
  \BibitemOpen
  \bibfield  {author} {\bibinfo {author} {\bibfnamefont {P.~W.}\ \bibnamefont
  {Milonni}},\ }\href@noop {} {\emph {\bibinfo {title} {The quantum vacuum: an
  introduction to quantum electrodynamics}}}\ (\bibinfo  {publisher} {Academic
  press},\ \bibinfo {year} {2013})\BibitemShut {NoStop}%
\bibitem [{\citenamefont {Mandel}\ and\ \citenamefont
  {Wolf}(1965)}]{MandelReview}%
  \BibitemOpen
  \bibfield  {author} {\bibinfo {author} {\bibfnamefont {L.}~\bibnamefont
  {Mandel}}\ and\ \bibinfo {author} {\bibfnamefont {E.}~\bibnamefont {Wolf}},\
  }\bibfield  {title} {\bibinfo {title} {Coherence properties of optical
  fields},\ }\href@noop {} {\bibfield  {journal} {\bibinfo  {journal} {Reviews
  of modern physics}\ }\textbf {\bibinfo {volume} {37}},\ \bibinfo {pages}
  {231} (\bibinfo {year} {1965})}\BibitemShut {NoStop}%
\bibitem [{\citenamefont {{Brown}}(1974)}]{HanburyBook}%
  \BibitemOpen
  \bibfield  {author} {\bibinfo {author} {\bibfnamefont {R.~H.}\ \bibnamefont
  {{Brown}}},\ }\href@noop {} {\emph {\bibinfo {title} {{The intensity
  interferometer; its application to astronomy}}}}\ (\bibinfo {year}
  {1974})\BibitemShut {NoStop}%
\bibitem [{\citenamefont {{Dravins, Dainis}}\ \emph {et~al.}(2015)\citenamefont
  {{Dravins, Dainis}}, \citenamefont {{Lagadec, Tiphaine}},\ and\ \citenamefont
  {{Nu\~nez, Paul D.}}}]{refId0}%
  \BibitemOpen
  \bibfield  {author} {\bibinfo {author} {\bibnamefont {{Dravins, Dainis}}},
  \bibinfo {author} {\bibnamefont {{Lagadec, Tiphaine}}},\ and\ \bibinfo
  {author} {\bibnamefont {{Nu\~nez, Paul D.}}},\ }\bibfield  {title} {\bibinfo
  {title} {Long-baseline optical intensity interferometry - laboratory
  demonstration of diffraction-limited imaging},\ }\href
  {https://doi.org/10.1051/0004-6361/201526334} {\bibfield  {journal} {\bibinfo
   {journal} {A\&A}\ }\textbf {\bibinfo {volume} {580}},\ \bibinfo {pages}
  {A99} (\bibinfo {year} {2015})}\BibitemShut {NoStop}%
\bibitem [{\citenamefont {{Classen}}\ \emph {et~al.}(2016)\citenamefont
  {{Classen}}, \citenamefont {{Waldmann}}, \citenamefont {{Giebel}},
  \citenamefont {{Schneider}}, \citenamefont {{Bhatti}}, \citenamefont
  {{Mehringer}},\ and\ \citenamefont {{von Zanthier}}}]{Superresolution}%
  \BibitemOpen
  \bibfield  {author} {\bibinfo {author} {\bibfnamefont {A.}~\bibnamefont
  {{Classen}}}, \bibinfo {author} {\bibfnamefont {F.}~\bibnamefont
  {{Waldmann}}}, \bibinfo {author} {\bibfnamefont {S.}~\bibnamefont
  {{Giebel}}}, \bibinfo {author} {\bibfnamefont {R.}~\bibnamefont
  {{Schneider}}}, \bibinfo {author} {\bibfnamefont {D.}~\bibnamefont
  {{Bhatti}}}, \bibinfo {author} {\bibfnamefont {T.}~\bibnamefont
  {{Mehringer}}},\ and\ \bibinfo {author} {\bibfnamefont {J.}~\bibnamefont
  {{von Zanthier}}},\ }\bibfield  {title} {\bibinfo {title} {{Superresolving
  Imaging of Arbitrary One-Dimensional Arrays of Thermal Light Sources Using
  Multiphoton Interference}},\ }\href
  {https://doi.org/10.1103/PhysRevLett.117.253601} {\bibfield  {journal}
  {\bibinfo  {journal} {\prl}\ }\textbf {\bibinfo {volume} {117}},\ \bibinfo
  {eid} {253601} (\bibinfo {year} {2016})},\ \Eprint
  {https://arxiv.org/abs/1608.03340} {arXiv:1608.03340 [quant-ph]} \BibitemShut
  {NoStop}%
\bibitem [{\citenamefont {Loudon}(2000)}]{Loudon}%
  \BibitemOpen
  \bibfield  {author} {\bibinfo {author} {\bibfnamefont {R.}~\bibnamefont
  {Loudon}},\ }\href@noop {} {\emph {\bibinfo {title} {The quantum theory of
  light}}}\ (\bibinfo  {publisher} {OUP Oxford},\ \bibinfo {year}
  {2000})\BibitemShut {NoStop}%
\bibitem [{\citenamefont {Mehringer}\ \emph {et~al.}(2018)\citenamefont
  {Mehringer}, \citenamefont {M{\"a}hrlein}, \citenamefont {von Zanthier},\
  and\ \citenamefont {Agarwal}}]{Mehringer2018}%
  \BibitemOpen
  \bibfield  {author} {\bibinfo {author} {\bibfnamefont {T.}~\bibnamefont
  {Mehringer}}, \bibinfo {author} {\bibfnamefont {S.}~\bibnamefont
  {M{\"a}hrlein}}, \bibinfo {author} {\bibfnamefont {J.}~\bibnamefont {von
  Zanthier}},\ and\ \bibinfo {author} {\bibfnamefont {G.~S.}\ \bibnamefont
  {Agarwal}},\ }\bibfield  {title} {\bibinfo {title} {Photon statistics as an
  interference phenomenon},\ }\href@noop {} {\bibfield  {journal} {\bibinfo
  {journal} {Optics letters}\ }\textbf {\bibinfo {volume} {43}},\ \bibinfo
  {pages} {2304} (\bibinfo {year} {2018})}\BibitemShut {NoStop}%
\end{thebibliography}%

\end{document}